# Tracing spatial confinement in semiconductor quantum dots by high-order harmonic generation

H. N. Gopalakrishna[1,2], R. Baruah[3,4], C. Hünecke[4], V. Korolev[1], M. Thümmler[4], A. Croy[4], M. Richter[4], F. Yahyaei[5], R. Hollinger[1], V. Shumakova[6], I. Uschmann[1,2], H. Marschner[1,2], M. Zürch[1,7,8], C. Reichardt[3], A. Undisz[9,10], J. Dellith[3], A. Pugžlys[6], A. Baltuška[6], C. Spielmann[1,2,11], U. Peschel[5,11], S. Gräfe[4,11,12], M. Wächtler[3,4,11], and D. Kartashov[1,11,*]

1. Institute of Optics and Quantum Electronics, Friedrich-Schiller University Jena, Max-Wien-Platz 1, 07743 Jena, Germany
2. Helmholtz-Institut Jena, Helmholtzweg 4, 07743 Jena, Germany
3. Leibniz Institute of Photonic Technology, Albert-Einstein-Straße 9 07745 Jena, Germany
4. Institute of Physical Chemistry, Friedrich Schiller University Jena, Helmholtzweg 4, 07743 Jena, Germany
5. Institute of Condensed Matter Theory and Solid-State Optics, Friedrich Schiller University Jena, Fröbelstieg 1, 07743 Jena, Germany
6. Institute for Photonics, Vienna University of Technology, Gußhausstrasse. 25-29, 1040 Vienna, Austria
7. Department of Chemistry, University of California, Berkeley, California 94720, USA
8. Materials Sciences Division, Lawrence Berkeley National Laboratory, Berkeley, California 94720, USA
9. Institute of Materials Science and Engineering, Chemnitz University of Technology, Erfenschlager Str. 73, 09125 Chemnitz, Germany
10. Otto Schott Institute of Materials Research, Friedrich Schiller University Jena, Löbdergraben 32, 07743 Jena, Germany
11. Abbe Center of Photonics, Albert-Einstein-Straße 6, 07745 Jena, Germany
12. Fraunhofer Institute for Applied Optics and Precision Engineering, Albert-Einstein-Str.7, 07745 Jena, Germany

Abstract

We report here on results of systematic experimental-theoretical investigation of high-order harmonic generation (HHG) in layers of CdSe semiconductor quantum dots of different sizes and a reference bulk CdSe thin film. We observe a strong decrease in the efficiency, up to complete suppression of HHG with energies of quanta above the bandgap for the smallest dots, whereas the intensity of below bandgap harmonics remains weakly affected by the dot size. In addition, it is observed that the ratio between suppression of above gap harmonics versus below gap harmonics increases with driving wavelength. These systematic investigations allow us to develop a simple physical picture explaining the observed suppression of the highest harmonics: the discretization of electronic energy levels seems to be not the predominant contribution to the observed suppression but rather the confined dot size itself, causing field-driven electrons to scatter off the dot's walls. The reduction in the dot size below the classical electron oscillatory radius and the corresponding scattering limits the maximum acceleration by the laser field. Moreover, this scattering leads to a chaotization of motion, causing dephasing and a loss of coherence, therefore suppressing the efficiency of the emission of highest-order harmonics. Our results demonstrate a new regime of intense laser-nanoscale solid interaction, intermediate between the bulk and single molecule response, and are crucial for nanophotonic platforms aiming control over high-order harmonic properties and efficiency.

# I. Introduction

After the first theoretical prediction [1] and experimental demonstration [2], high-order harmonic generation (HHG) in bulk solids became a rapidly growing research field in ultrafast strong-field physics. Numerous experiments have shown that, due to the nonlocal nature of the strong field-driven nonlinearity of the carrier's quantum dynamics, HHG in crystalline solids has great potential as an ultrafast spectroscopy probe, sensitive to crystal symmetry and structure [3, 4], electronic band structure [5-7] etc. with a sub-cycle temporal and picometer spatial resolution [8, 9].

The physics of HHG in bulk solids is well understood. After electron excitation from the valence to the conduction band by the strong laser field, high-order harmonics are generated by two mechanisms: the intraband electron-hole currents in the conduction and valence bands (nonlinear Bloch current, a consequence of the non-parabolicity of the bands), and nonlinear polarization between the field-driven electrons and holes during their phase-locked motion in the corresponding bands. The first mechanism contributes mostly to harmonics with energies of quanta below the bandgap, whereas the second mechanism dominates the emission spectrum above the bandgap and ultimately requires coherent electron-hole motion [10].

The question that we address here is: How is the HHG process impacted by confinement, when crystallite's size is reduced down to the nanoscale? Spatial confinement leads to break down of basic concepts used in solid state physics and have multiple effects on the carrier dynamics in solids. A spatial extent of the system of only a few unit cells leads a) to a discretized energy structure, putting under question the concept of bands itself; b) to effective masses becoming less well defined, because they depend not only on the band, but also on the actual position in the coordinate space within the nanocrystallite; c) to a reduction of the relative contribution of the volume towards an increased role of surface states; and d) to scattering from the potential barrier at the crystallite's surface ("off-the-wall" scattering) when the crystallite's size is comparable or less than the amplitude of the oscillatory motion of the electron wavepacket, excited and driven by the laser field. At the same time, nanocrystallites, still being composed of hundreds or even thousands of atoms, are still far from the atomic or molecular scales. Therefore, the question is how can we trace the impact of confinement on HHG and which mechanisms contribute to HHG most in strongly confined systems?

Very recently, the first experimental results on HHG in CdSe quantum dots of different size and fixed 3.5 µm laser wavelength have been reported [11]. In these experiments, an abrupt drop in the harmonic intensity for quantum dots with a diameter less than 3 nm has been observed. It has been suggested that this effect originates from the discrete electronic energy structure in the conduction band of small dots that leads to a suppression of intraband transitions and by that, a suppression of HHG [11]. However, several aspects in these results remain puzzling: for example, calculations of the energy level structure for the dots with sizes up to 2 nm show that the energy spacing between the lowest unoccupied levels is still much smaller than the energy of the laser quanta used in the experiments in [11]. Therefore, it is not clear why the discretization would lead to a suppression of intraband harmonics. To shed light on this phenomenon, we present here a more detailed experimental and theoretical study in a broad parameter range.

Here, we investigate experimentally HHG in layers of CdSe semiconductor quantum dots (QD) of different diameter as a function of the laser intensity, ellipticity, and wavelength. We likewise see that above bandgap harmonics show an abrupt and pronounced drop in intensity for dot diameters below 3 nm, in agreement with [11]. Moreover, we demonstrate that this effect becomes especially pronounced for longer laser wavelengths. However, harmonics with energies near the bandgap energy, that have comparable contribution of intraband and interband generation mechanisms [10], are not sensitive to change in the dot diameter. Also, we investigate the dependence of harmonic yield on the pumping intensity and ellipticity as a function of dot's diameter. We show that, whereas the ellipticity dependence is almost the same in the reference bulk and dots of all sizes, the intensity dependence demonstrates a gradual increase in the nonlinearity of HHG with reducing the dot's size.

Our experimental results, supported by numerical simulations, suggest that, despite the visible discretization of the electronic energy structure, this seems to play a minor role, as the energy spacing is still much smaller than the driving frequency for all cases. Instead, the major effect on HHG efficiency can be explained within a simple classical picture: for dot sizes below the classical oscillatory radius of the electron in the laser field, scattering at the dots' walls leads to an out-of-phase dynamics relative to the laser optical cycles, and, thus, to a reduced acceleration. This, in turn, prevents the generation of the highest harmonics.

The manuscript is organized as follows. The experimental setup, sample design and characterization are described in section II and the Supplemental Material. The results of experimental measurements are discussed in section III. The results of numerical simulations are presented in section IV, followed by discussions and conclusions.

## II. Experimental setup and sample characterization

The experiments were carried out using a femtosecond optical parametric amplifier system (Light Conversion TOPAS Prime) pumped by a 35 fs laser source at 0.8 µm wavelength (Coherent, Astrella), followed by a difference frequency generation stage, providing 70-100 fs pulses tunable in the spectral range 3-16 µm. The temporal shape of the pulses was characterized via Frequency-Resolved Optical Gating (FROG) based on second harmonic generation (SHG), whereas the focal intensity distribution was characterized using a mid-IR CCD camera. The experimental setup is depicted in Figure 1. The mid-IR laser beam was focused by a $f$ = 75 mm lens at normal incidence onto a quantum dot layer, deposited on a substrate, in a spot of 85 µm diameter ($e^{-2}$ intensity level). The high-order harmonic radiation transmitted through the substrate was collimated by a $CaF_2$ lens and then focused on the entrance slit of a spectrometer (Kymera-328i) equipped by a cooled CCD camera as a detector. The spectra were measured in the spectral range 200 – 1100 nm limited by the spectrometer and the detector. We also employed an on-site microscopy consisting of a white light source and a 50X microscope objective for optical diagnostic of the quantum dot layers before and after interaction with laser pulses. The microscope also allowed alignment of the HHG measurements in the vicinity of a marked area of the layer, thoroughly characterized with different diagnostic methods as described in what follows.

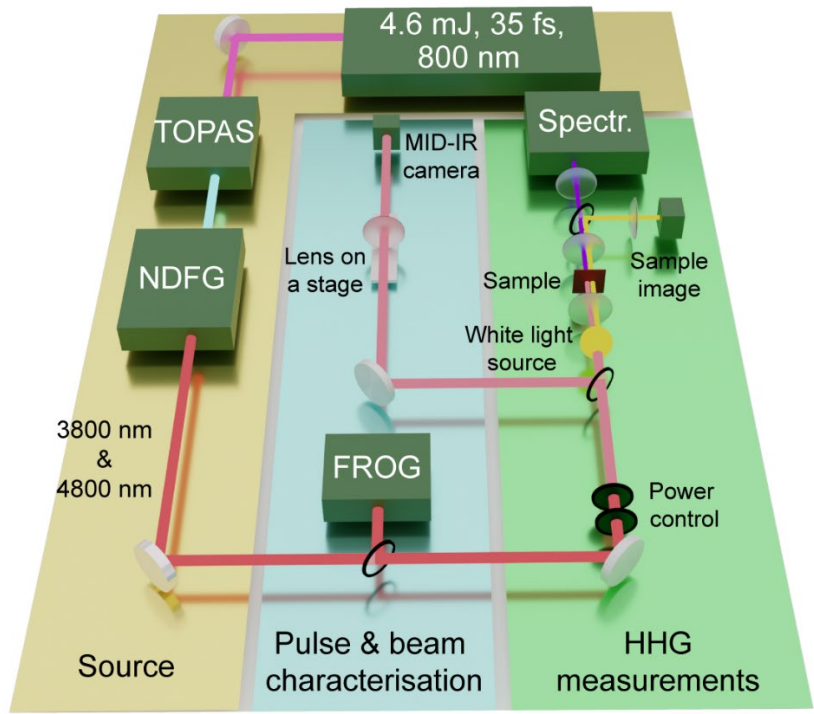

FIG. 1 Experimental setup: NDFG – non-colinear difference frequency generation module, providing tunable mid-IR pulses by DFG between the signal and idler waves; FROG – Frequency-Resolved Optical Gating apparatus for pulse characterization. The lens on stage with a mid-IR CCD camera was used for focus beam profile characterization. A white-light source with an 50x removable infinity-corrected objective (not shown), a lens and a CCD camera were used as an imaging system for the on-site sample diagnostic.

Colloidal CdSe quantum dots (QD) were synthesized following established protocols via the hot injection method (for details see Supplemental Material) [12, 13]. The surface of the QDs was covered by trioctylphosphine oxide (TOPO) which renders the QDs dispersible in non-polar solvents, e.g. toluene. By variation of injection temperature and growth time, the size of the QDs was controlled and varied in the range between 2 and 9 nm. As expected, due to quantum confinement, the position of the excitonic transitions in the absorption and photoluminescence spectra of the particles shift in dependence on size (see Supplemental Material Figure S1 and S2) [14]. The band gap was determined from a Tauc plot [15] and the position of the band gap photoluminescence peak. Both methods resulted in similar values (Table 1). The size of the nanocrystals was determined from the spectral position of the first excitonic transition in the absorption spectra by the method described in [16] and compared to the results obtained from transmission electron microscopy (TEM) images (see Figure 2 and SM). A typical example of the statistical variations of dots' diameter $d$ is shown in Figure 2(b), suggesting that full-width half-maximum (FWHM) spread of the size is within 10% (for a documentation of the

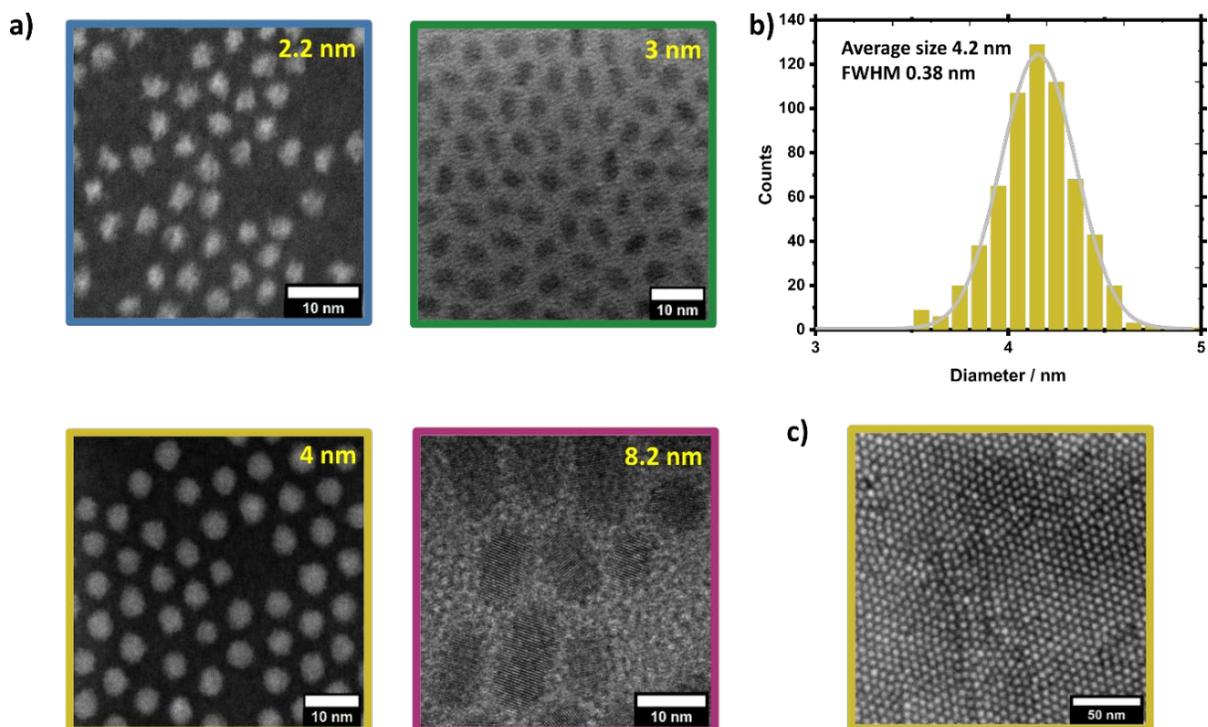

FIG. 2 a) TEM-images of quantum dots of different sizes used in this investigation. Acquisition was either carried out in scanning TEM mode using a high-angle annular detector or in TEM bright-field mode. b) Statistical size distribution for 4 nm QDs. The analysis of the TEM images results in a mean diameter of 4.2 nm, which is in good agreement with the diameter of 4 nm determined from the position of the first excitonic transition in the absorption spectra. c) SEM image of a layer of 4.2 nm dots deposited at a Si substrate.

size distribution analysis for all samples, see Supplemental Material). The crystal structure was determined using X-ray diffraction (XRD) to be wurtzite (see Supplemental Material, Figure S3, S4). Note, that all quantum dots employed in our experiments satisfy strong confinement criteria, meaning that the dots' diameter is smaller than exciton size in the bulk, which is for CdSe roughly 11 nm [17].

QDs dispersed in toluene were drop-casted on a 500 µm thick sapphire substrate to form a thin well-ordered QD layer after solvent evaporation (Figure 2(c)). We targeted a layer thickness in the µm range as was confirmed by atomic force microscopy (AFM) (Table 1 and SM). The photoluminescence peak positions determined for QD layers deposited on sapphire are identical

TABLE I. Characteristics of quantum dot layers

| Dot diameter $d$ from peak position, nm | Dot diameter $d$ from TEM, nm | Bandgap from Tauc plot, eV | Bandgap from emission peak, eV | Layer thickness (µm) |
|---|---|---|---|---|
| 2.2 | 2.6 | 2.48 | 2.48 | 2.4 |
| 3 | 3.4 | 2.18 | 2.21 | 3.8 |
| 4 | 4.2 | 2.05 | 2.08 | 4.5 |
| 8.2 | 7.7[a] | 1.89 | 1.89 | 2.8 |

a) The 8.2 nm QDs are not exactly spherical and show an aspect ratio of long and short axes of 1.7 :1.

to the spectra in solution (see Supplemental Material, Figure S1), i.e., no changes in spectral position of PL peaks are observed. This confirms that there is no significant electronic interaction between the neighboring QDs after deposition, and the QDs can be regarded as isolated particles due to the presence of bulky TOPO ligand. Thus, the dots in the layers can be considered as independent emitters of high-order harmonics.

Finally, as a reference bulk sample, we prepared a 140 nm thick polycrystalline CdSe film consisting of randomly oriented wurtzite crystallites with a diameter of 52 nm and a bandgap of 1.64 eV. The film was grown on a sapphire substrate by the chemical bath deposition method [18] with subsequent vacuum annealing (detailed description see Supplemental Material).

## III. Experimental results

The spectra were measured by irradiating quantum dots of different size with 100 fs laser pulses at a fixed laser intensity of 1.2 TW/cm$^2$, at two different wavelengths 3.72 µm and 4.74 µm, are shown in Figure 3(a),(b). All spectra are corrected for the spectrometer sensitivity and acquisition time. Thus, the spectra for a fixed laser wavelength λ but different dot size can be directly compared. For both laser wavelengths, we observe significantly faster decay in amplitudes of harmonics with increasing harmonic order in quantum dots in comparison to the reference bulk film. Moreover, we observe an abrupt drop in the intensity of harmonics with energies of quanta above the bandgap for dots with a diameter below $d$ =3 nm, in agreement with the results in [11].

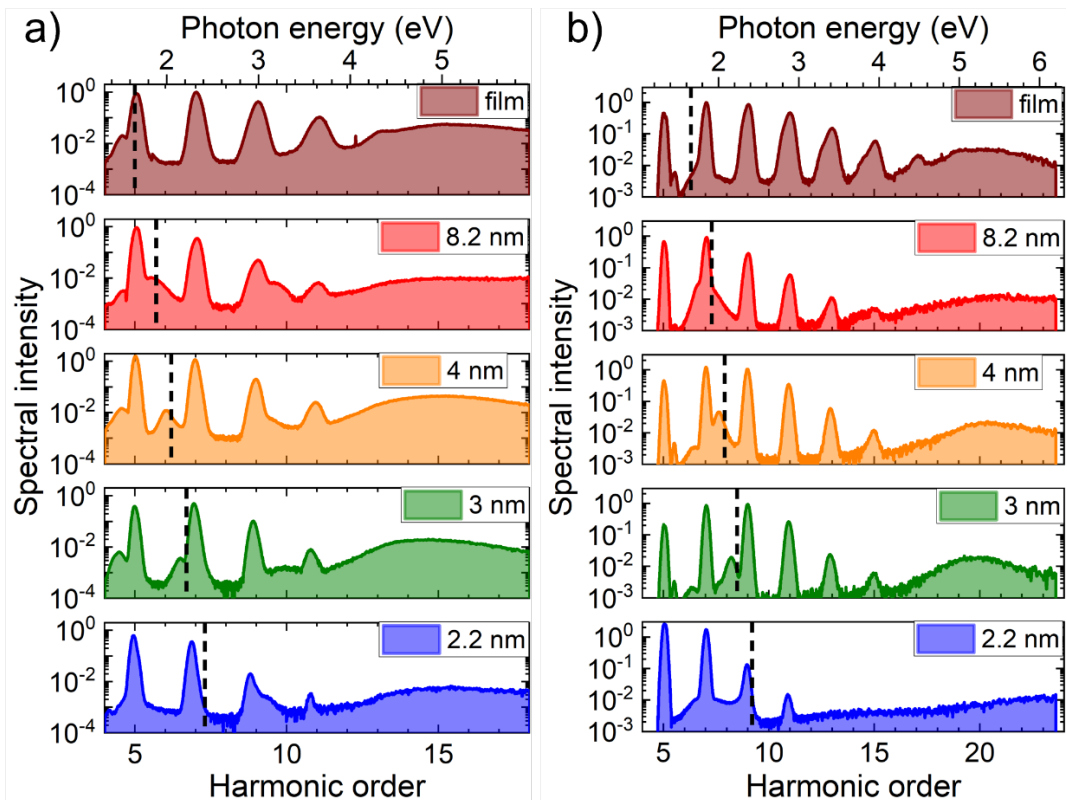

FIG. 3 Measured HHG spectra in different size QDs for a) 3.72 µm and b) 4.74 µm laser wavelength at a fixed intensity. The dashed line marks the bandgap determined from photoemission/photo absorption measurements.

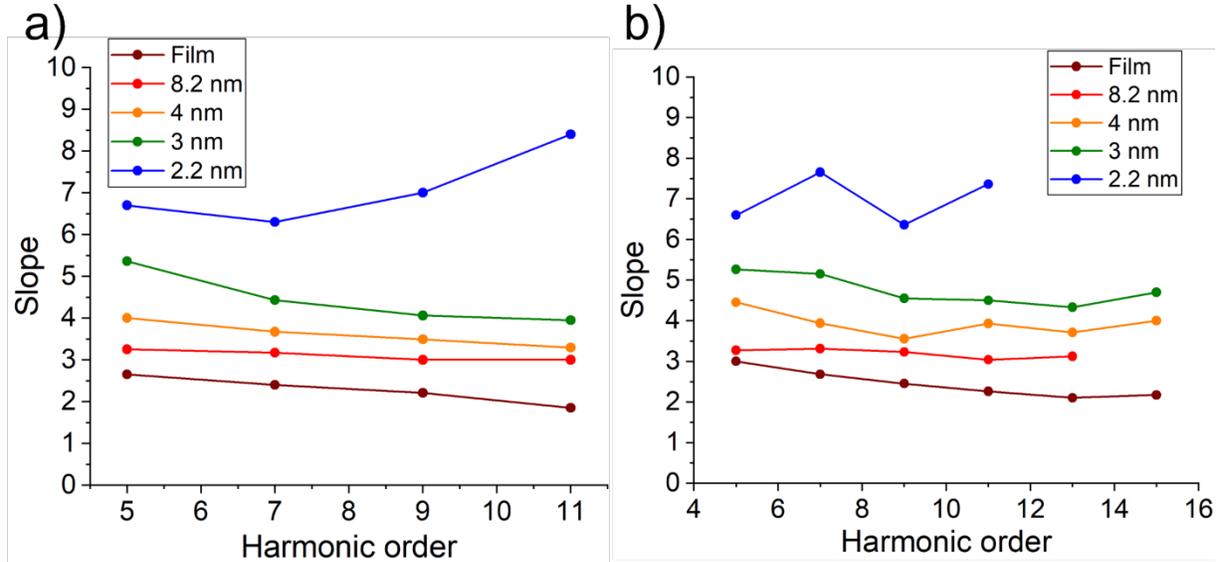

FIG. 4 The slope of the laser intensity dependence of harmonic's yield as a function of harmonic order in the range of laser intensities 0.4 - 1 TW/cm$^2$ for a) 3.75 μm and b) for 4.74 μm laser wavelength

This effect is especially pronounced for the longer λ = 4.74 μm wavelength, where essentially all harmonics above the bandgap are strongly suppressed. At the same time, the efficiency of generation of the below and near the bandgap harmonics (5$^{th}$ and 7$^{th}$ for λ=3.72 μm and 5$^{th}$-9$^{th}$ for λ=4.74 μm) remains almost unaffected by the dot's diameter.

We further investigated the dependence of the harmonic yields on the laser intensity for both wavelengths. We found that, in the range of laser intensities 0.4 - 1 TW/cm$^2$ used in the experiments, the intensity dependence of the yield for the individual harmonics can be very well fitted in a double logarithmic scale by a linear function (see Supplemental Material for details). The slope of the corresponding linear fit, i.e. the power index, as a function of harmonic's order and different dot's size is shown in Figure 4. For the film and large diameter dots, the power index is noticeably lower than the harmonic order, confirming the non-perturbative character of the nonlinearity. Also, the power index slightly decreases or remains approximately constant as a function of harmonic order (for a fixed dot diameter). However, for all harmonic orders the power index is higher for smaller QDs. Moreover, for the smallest QD with d = 2.2 nm the power index for the below bandgap harmonics (5$^{th}$ for λ = 3.75 μm and 5$^{th}$ and 7$^{th}$ for λ = 4.74 μm) is even higher than the corresponding harmonic order. This is unexpected, because one would expect that restriction in the amplitude of the electron wavepacket motion within the nanocrystal should lead to more like perturbative character of nonlinearities, i.e. the power index should remain below or be close to the harmonic order.

Finally, we measured the polarization dependence of HHG in the reference polycrystalline film and quantum dots of different sizes when changing the laser polarization from linear to circular. We analyzed the dependence of individual harmonic yield on the laser polarization ellipticity and the results for the full width at half-maximum (FWHM) in this dependence are shown in Figure 5 (the examples of the ellipticity dependence are shown in Supplemental Material). The qualitative behavior of the FWHM in the ellipticity dependence as a function of harmonic's order is the same in the film and all dots, dropping for below-near bandgap harmonics and then saturating or even slightly increasing for harmonics above the ninth. This slight increase in the width of the ellipticity dependence might be due to the random orientation of crystallites in the film and dots layers. It has been shown experimentally and theoretically before that ellipticity dependence of HHG in solids is more complicated than in gases, and that the efficiency of higher-orders harmonic generation might be increased depending on the orientation of crystal symmetry axes relative to the polarization ellipse [19], i.e. can have maximum at a certain ellipticity. To the best of our knowledge, these are first measurements of ellipticity dependence in QD's as a function of the dot's size. Similar behavior of this dependence for all dots and the bulk suggests, that the mechanism of HHG is the same in all these samples.

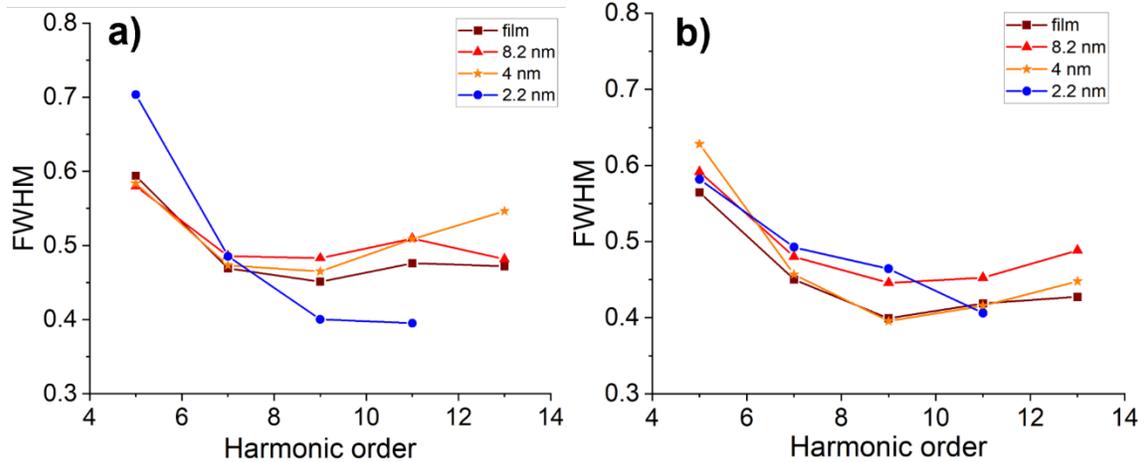

FIG. 5 FWHM of ellipticity dependence in HHG for a) 3.75 μm and b) for 4.74 μm laser wavelength and different dot diameter. The laser intensity is fixed at 0.5 TW/cm$^2$.

Our main experimental observations can be summarized as follows: (1) The intensity of the above bandgap harmonics drops abruptly when the dot diameter $d$ is reduced below 3 nm, whereas below and near bandgap harmonics remain largely unaffected; (2) the power index in the dependence of harmonic's yield on the laser intensity gradually increases with reduction in the dot size but its value is weakly dependent on harmonic's order. For the smallest $d$ = 2.2 nm dots the power index for below bandgap harmonics exceeds the corresponding harmonic order; (3) the bulk semiconductor and QD's of all sizes show a very similar dependence of the harmonic yield on the laser polarization ellipticity.

## IV. Numerical simulations

We first calculated the static spatial and electronic structure of the quantum dots. Due to the system's size, we calculated the electronic energy structure using the tight-binding model GFN1-xTB as implemented in the XTB program package [20]. The structure and energies for different sizes of quantum dots, ranging from small $Cd_2Se_2$ molecular-type units up to CdSe clusters with diameters of $d \approx 4$ nm were calculated. The corresponding results for the d ≈ 2.2 nm dot ($Cd_{136}Se_{136}$) and $Cd_2Se_2$ molecule, equivalent to one unit cell in the bulk crystal, are shown in Figure 6. First, it should be noted that the spatial structure of small quantum dots is different from the crystalline structure in the bulk. In Fig.6a the arrangements of Cd (white spheres) and Se (yellow spheres) atoms in 2.2 nm size dot is shown for the case as it would be in a cut from a wurtzite bulk crystal and for the case of energetically optimized geometry. The crystal structure is more compact than the optimized dot structure, as it is shown in histograms of interatomic distances between Cd and Se atoms for both structures in Figure 6(a). The electronic energy structure for the optimized geometry 2.2 nm dot is shown in Figure 6(b). As follows from Figure 6(b), with changing the size, not only the precise value of the bandgap (or, for the molecule, HOMO-LUMO gap, Fig.6c) changes. Also, the energy gap between the individual occupied and unoccupied states changes substantially, as expected: the larger the dot, the closer the energy levels (of similar electronic character) become, forming, in the limit of an infinite, periodic system, a band. For the dots of d≈2.2 nm diameter, the discrete structure of electronic states is well visible (Figure 6(b)). However, the energy level spacing is still substantially smaller than the energy of the laser photons ($\hbar\omega$=0.26 eV for $\lambda$= 4.74 μm laser), see inset in Figure 6(b). This suggests that the discretization of the electronic energy structure does not play a significant role for the suppression of harmonic radiation even for the smallest 2.2 nm diameter dots used in the experiments.

To simulate HHG in the bulk reference and quantum dots, we employed first ab-initio approach based on real-time real-space time-dependent density functional calculations (rt-TDDFT) using the OCTOPUS code (for details, see SM) [21]. However, these calculations are computationally extremely challenging. The largest dot, that we could simulate within a reasonable calculation time, consisted of 64 atoms dots ($Cd_{32}Se_{32}$), corresponding to optimized

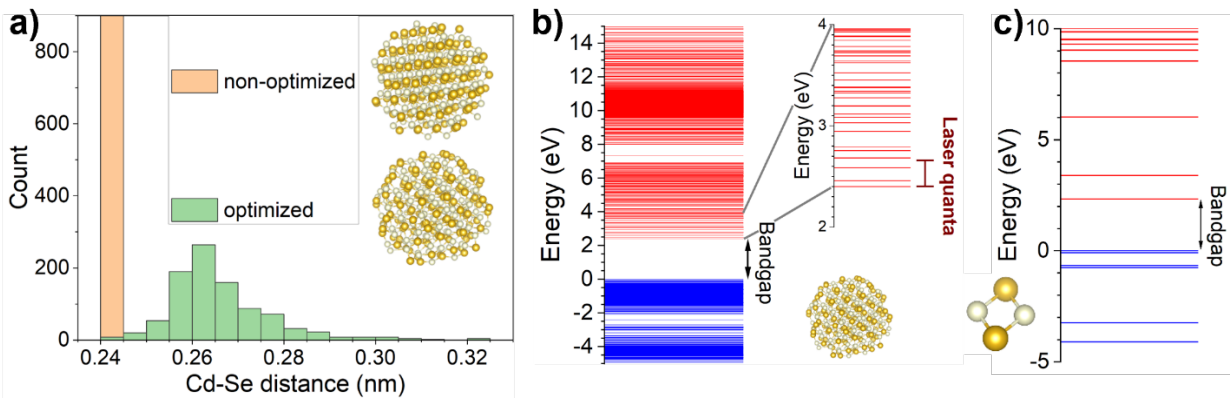

FIG. 6 a) Visual representation of 272 atom $Cd_{136}Se_{136}$ quantum dot (diameter 2.16 nm) with spatial structure as it would be as a cut from bulk (upper image) and energetically optimized structure (lower image). The histograms show the distribution of interatomic distances between Cd and Se atoms; b) electronic energy structure of $Cd_{136}Se_{136}$ quantum dot with optimized geometry and c) $Cd_2Se_2$ molecule. The blue levels are fully occupied and would form in the bulk valence band(s), whereas the red levels show unoccupied states and would form conduction bands. The inset in b) shows zoomed bottom of the conduction band. The bar shows the scale of the laser quanta energy 0.26 eV, corresponding to the 4.72 μm wavelength.

geometry of d ≈ 1.5 nm, requires >270 kCPU hours on a supercomputer cluster for one single set of laser parameters. For the bulk, a single calculation (without orientation averaging) requires >100 kCPU hours. Thus, these calculations confirm a strong suppression of harmonic intensities for d ≈ 1.5 nm diameter dots in comparison to the bulk, but do not allow a systematic study of dots with varying diameters. The details of these calculations are given in Supplemental Material.

To develop a different model, we refer to the experimentally observed difference in the behavior of below and above bandgap harmonics with reducing dot's size. We remind that the bandgap energy separates two mechanisms of harmonic generation – the nonlinear intraband current dominating harmonics below the bandgap and the interband recombination/polarization that generates harmonics above the bandgap [10]. As both intra- and interband currents are proportional to the carrier density, the observed weak sensitivity of near bandgap harmonics to dot's diameter and sharp drop for above bandgap harmonics for small dots suggest that this difference does not originates from a strongly reduced excitation rate for small dots, as has been suggested in [11] and explained by the discretization of the electronic energy structure. Instead, another mechanism, selectively affecting the carrier acceleration by the laser field and recombination, should be at play. Also, the qualitatively and quantitatively nearly identical dependence of the harmonic yield on the ellipticity of the laser polarization, suggests that the mechanisms of HHG should be the same for the film and QD's of all sizes.

Therefore, we applied a commonly used simple model to analyze the origin of the experimentally observed confinement effects in HHG. Shortly, we numerically solve the one-dimensional time-dependent Schrödinger equation (TDSE) for a single-active electron in a box of finite size to investigate the laser field-driven quantum dynamics of electrons in the dots. The dot is modeled by the potential $V(x) = \begin{cases} -V_0, & |x| \leq d/2 \\ -\dfrac{q}{\sqrt{(|x|-d/2)^2 + \dfrac{q^2}{V_0^2}}}, & |x| > d/2 \end{cases}$. Here, the amplitude $V_0$ defines the specific work of the dot, the wall charge $q$ determines the steepness of the wall potential, and $d$ is the dot diameter. In the simulations, we choose $V_0$=0.24 a.u. (6.6 eV) to mimic the work function of CdSe [22] and a wall charge $q$=1. We note that the precise choice of the wall charge does not influence the results substantially. For the chosen parameters, the energy gap between the stationary ground and the first excited states in the dots of 2.2 nm, 3 nm and 4 nm diameter is 0.154 eV, 0.092 eV and 0.056 eV correspondingly. These energies are less than the energy of quanta for the laser wavelengths 3.75 μm and 4.75 μm used in simulations. Therefore, the particle in the box model has the following important differences from calculations based on the bulk band structure: 1) as it is a single-particle model, many-body contributions (such as electron-hole dynamics etc.) are intrinsically not accounted for; 2) despite the very dense energy spacing in the model, there are neither quantitatively correct valence energy levels, nor a proper energy gap. In our model, immediately with the onset of the laser field, an electron wavepacket is launched and start to move across the box ("inside the quantum dot", mimicking the current contribution to HHG; 3) The electron wavepacket driven inside the potential well may bounce off the dots' walls. This leads to dynamics which is not periodic with respect to the laser field, therefore, the emitted spectrum does not feature pronounced harmonics but rather a continuum structure; 4) as the potential inside the well is flat, harmonics are generated only when an electron wavepacket is accelerated near the walls, in the region where the dipole acceleration $\ddot{d} = \int_{-\infty}^{\infty} |\Psi|^2 \frac{\partial V}{\partial x} dx$ is non zero. We note parenthetically that a model potential which includes an atomic like structure, does not change the results noticeably. We therefore restrict ourselves to the simple model to extract the phenomena more clearly. Despite of its shortcomings, our simple model still gives us an idea about the motion of a laser driven charged particle under the influence of confinement. In fact, the main features of the particle dynamics, as will be shown later, can even be well described classically.

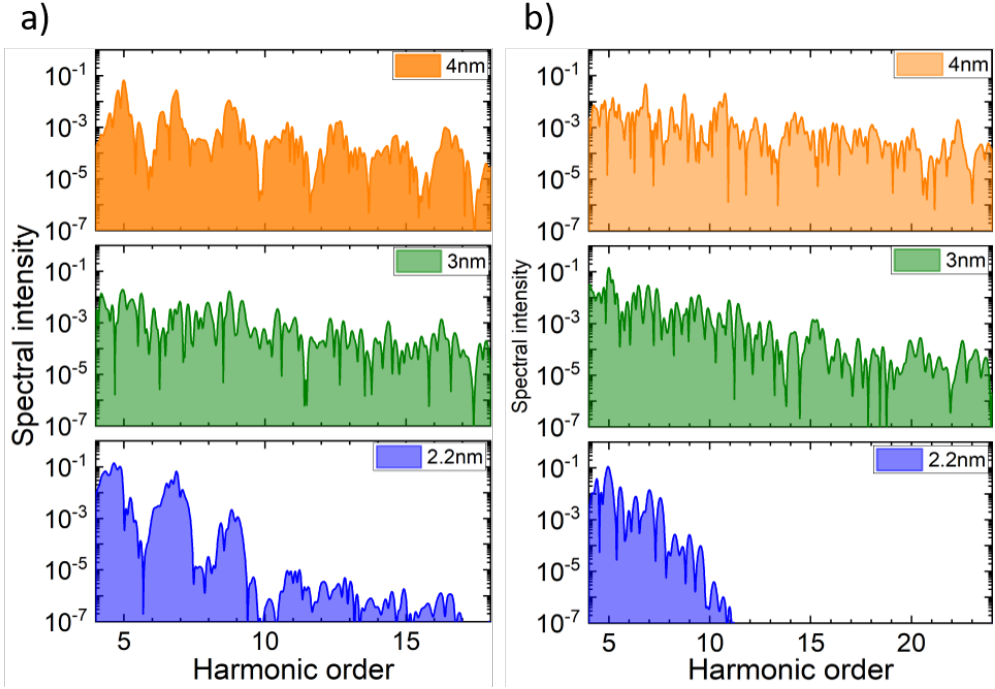

FIG. 7 Simulated dipole acceleration spectra in QD's of different size for a) 3.75 µm and b) 4.75 µm laser wavelengths. The irregular structure of the spectrum is the consequence of scattering at the walls, see in the text.

The calculated spectra of the dipole acceleration for the laser intensity 0.3 TW/cm$^2$, pulse duration 75 fs FWHM (sin$^2$ envelope), laser wavelengths λ= 3.75 µm and 4.75 µm and different dot sizes are shown in Figure 7. The results of the quantum dynamical simulations reproduce very well the experimentally observed effect of suppression of highest order harmonics when reducing the dot's size. Also, they confirm that the harmonic suppression becomes more pronounced with increasing the laser wavelength. For the chosen parameters, ionization (expressed in terms of loss of norm of the wavefunctions caused by parts of the electronic wavefunction leaving the grid) remains throughout negligible. Therefore, we can conclude that harmonics are generated due to wavapacket dynamics within the potential well.

As mentioned above, the systems' dynamics are even well described classically. Thus, to get further insight, we complemented the TDSE quantum calculations by a one-dimensional Monte-Carlo trajectory analysis, based on $10^6$ electron trajectories weighted by Wigner probability distribution over the initial coordinate and momentum distribution of the quantum mechanically calculated ground state in the model potential, see Supplementary Material for details. We used for these calculations the same laser pulse parameters as were used in TDSE calculations. In Figure 8, the dynamics of some selected (neighboring) trajectories are displayed. We see very clearly that for larger dots, the dynamics directly follows the laser oscillations with an oscillatory radius of $r_{osc} = E_0/\omega^2$ (in atomic units), while in smaller dots, when $d < r_{osc}$, trajectories bounce off the walls. This, in turn, has two consequences: (i) the scattering off the walls leads to a kind of chaotization of dynamics, which, in a quantum picture, would correspond to a loss of coherence; (ii) as the trajectories do not follow the full length in the oscillatory radius $r_{osc}$, the electron cannot acquire in full extend the corresponding (ponderomotive) energy. To verify this, we calculated the kinetic energy of each trajectory within the time window when the electron is moving in the region where in our model harmonic emission would occur (near the walls); we define this region as the interval from dot's radius till the point, where the gradient of the potential drops to 1% from its maximum value $\frac{2}{3\sqrt{3}}\left|\frac{V_0}{q}\right|$ located at the points $|x| = \frac{d}{2} + \left|\frac{q}{\sqrt{2V_0}}\right|$. The results of calculations for the same laser pulse parameters, as were used in TDSE simulations, are shown in Figure 9. Two important conclusions follow from the analysis of Figure 9. First, there is a clear reduction in the maximum kinetic energy acquired by electrons for dots with a diameter d< 3 nm for both driving laser wavelengths. This reduction is larger for the longer wavelength λ, in agreement with the TDSE simulations and the simple classical considerations, i.e. when $d < r_{osc}$, (see Figure 7). Second, there is clear chaotization of the electron trajectories' dynamics caused by the scattering off the wall. Without the scattering, all electron trajectories would show regular oscillations with half of the optical cycle periodicity (compared to Figure 8), reflected in regular oscillations in the kinetic energy (can be seen in Figure 9 for relatively low energies).

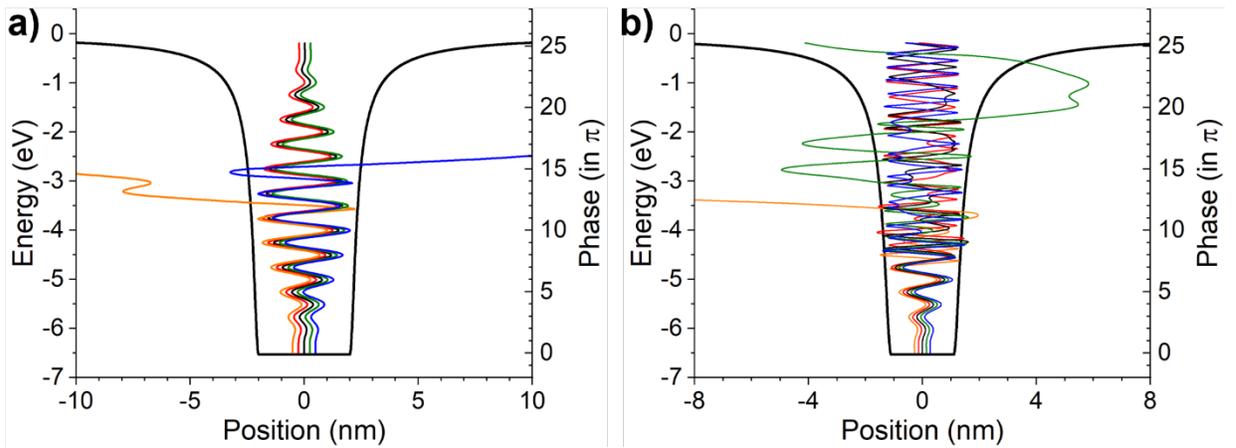

FIG. 8 Dynamics of selected trajectories with zero initial velocity and different initial positions within the well for a) 4 nm diameter dots and b) 2 nm diameter dots. The phase of the laser field oscillations is defined as $\varphi = \omega t/\pi$ where $\omega$ is the laser frequency. The trajectories leaving the well corresponds to particles, stochastically heated to energies above the work function $V_0$.

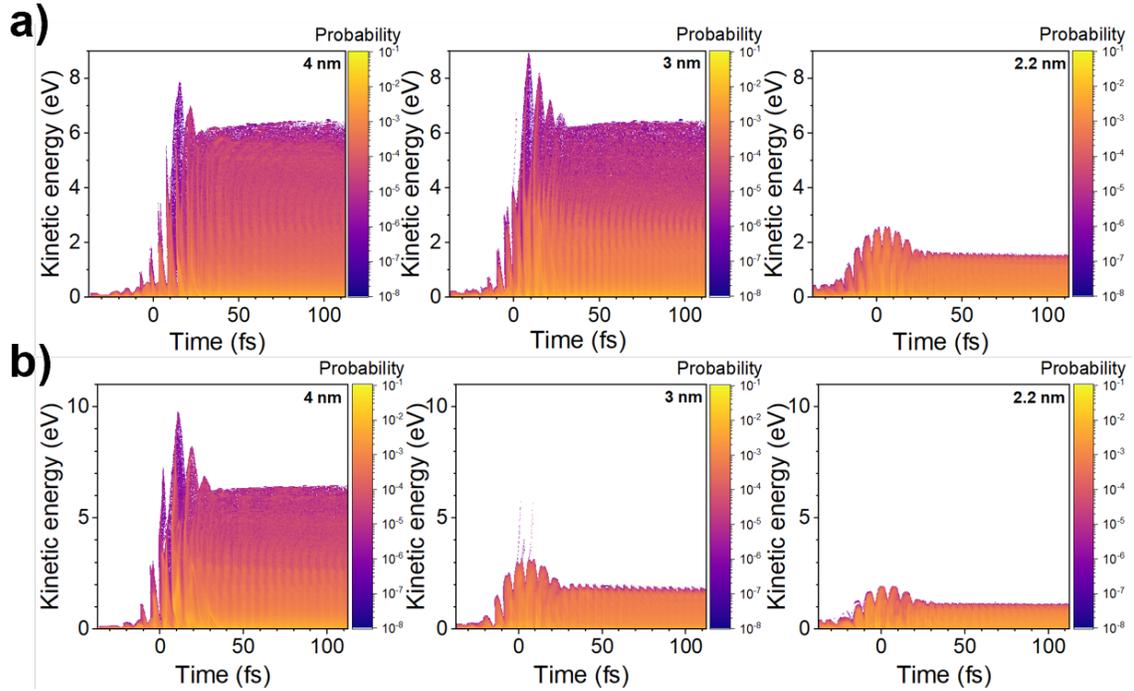

FIG. 9 Kinetic energy at different electron trajectories, moving at a given moment of time within the emission region, for different diameter of the dots and driven by a) 3.75 µm and b) 4.75 µm laser wavelength.

The scattering leads to a dephasing between the trajectories and washing out of regular oscillation structure in the kinetic energy (for energies roughly above 4 eV in Figure 9).

Thus, the trajectory analysis suggests that the origin of the suppression in harmonic yield for highest harmonic orders in small dots is caused by two contributions: (i) the small dots sizes prevent full oscillatory dynamics of the electron inside the dot, leading to a reduced acquisition of kinetic energy and, thus, prohibits emission of highest harmonics; (ii) the scattering off the dot's wall leads to a chaotization of the dynamics, which corresponds to dephasing and decoherence in the quantum description, and causes an additional suppression of emission of the highest harmonic-orders. These effects become important when the dot size $d$ is less than the oscillatory radius $r_{osc}$ of electron motion, thus explaining the experimentally observed scaling with the driving wavelength and intensity.

Finally, we tried to reproduce in frame of the particle-in-a box model the increased nonlinearity in the intensity scaling of harmonics. For this we introduced a position dependent effective mass of electron, such that at the interface the electron's effective mass changes from that of the bulk (for wurtzite CdSe $m_{eff} = 0.13\ m_0$) to that of free space $m_0$. Also, we included the effect of the optical field enhancement, which leads to an increase in the field amplitude outside the QD up to factor 2.2. Both effects cause an abrupt change of acceleration and consequently an increase of HHG resulting in a stronger power dependence as soon as the optical field forces the electron to leave the QD. However, we were not able to reproduce the increase in the power index for 5[th] harmonic up to the value 6.7, retrieved from the experimental data.

As stated above, this very simple particle-in-a box model cannot describe the coherent electron-hole dynamics in the conduction and valence bands giving rise to the emission of harmonics larger than the bandgap. However, this interband recombination mechanism of HHG is very sensitive to dephasing of the electron-hole motion. Scattering of the electron wavepacket off the dot's boundary will ultimately lead to a dephasing of the electron-hole motion and a momentum mismatch. Therefore, we can expect that this dephasing of the electron-hole motion, induced by the off-wall scattering, will also lead to a significant drop in the efficiency of HHG above the bandgap.

## V. Conclusions

In conclusion, we experimentally investigated the confinement sensitivity of HHG in layers of semiconductor quantum dots as a function of dot's diameter, laser intensity, polarization and wavelength. We have shown that a reduction in size of quantum dots $d$ below the oscillatory radius $r_{osc}$ of the electron wavepacket motion results in strong suppression of highest-order harmonics with energies of quanta above the bandgap. Since these harmonics are predominantly generated by the electron-hole recombination mechanism, an analog of Corkum's three-step HHG mechanism in gases [23, 24], we suggest that this suppression is due to inefficient acceleration of the electron wavepacket, spatially confined within dot's. Numerical simulations, based on rt-TDDFT, and a simple quantum dynamical model TDSE for a particle in a box, confirm the strong suppression of harmonic emission. An accompanying classical trajectory analysis, based on Monte Carlo simulations, clearly shows that for dots with a diameter $d < r_{osc}$, trajectories bounce off the dot's walls, leading (i) to a substantial reduction of acquired ponderomotive energy and (ii) a chaotization and randomization of motion, leading to a loss of coherence. These two effects are responsible for the observed suppression of the highest-order harmonics. We believe that our experimental results demonstrate a new regime of extreme nonlinear optics in nanoscale solids, an intermediate between the bulk solid behavior and single molecule response. They mark a border set by strong quantum confinement in minimum size of nanostructuring in solids, that is crucial for the development of nanophotonic elements, metasurface and topological insulator structures for control and enhancement of HHG efficiency [25-27].

## Acknowledgement

C.H., M. T., V.K., R. B., F. Y., D. K., M. W., S. G. and U. P. gratefully acknowledge financial support by the German Research Foundation DFG under Collaborative Research Center SFB 1375 "NOA", Projects A1 and C4.  S. G. and M. R. gratefully acknowledge financial support by ERC within Consolidator Grant "QUEM-CHEM". M. R., C. H. and A. C. gratefully acknowledge the Supercomputer Cluster PC2 in Padeborn for grunting the calculation time. D. K., M. W. and C. R. gratefully acknowledge financial support by State of Thuringia within ProExcellence Initiative APC2020. M. W. gratefully acknowledge financial support by Fonds der Chemischen Industrie (FCI). A. B. and A. P. gratefully acknowledge financial support by the Austrian Science Fund (FWF) under Project I 4566. C. S. and M. Z. gratefully acknowledge financial support by Federal Ministry of Education and Research (BMBF) under "Make our Planet Great Again - German Research Initiative" (Grant No. 57427209 "QUESTforENERGY") implemented by DAAD. A. U. gratefully acknowledge financial support by the German Research Foundation within the infrastructure grant 390918228 – INST 275/391-1. The authors thank Dr. Marco Diegel for performing XRD measurements, Prof. Misha Ivanov for fruitful discussions and Dr. I. Gonoskov for discussions about 1 1D-TDSE dot calculations.

## References

[1] D. Golde, T. Meier, S. W. Koch, "High harmonics generated in semiconductor nanostructures by the coupled dynamics of optical inter- and intraband excitations," Phys. Rev. B **77,** 075330 (2008).

[2] S. Ghimire, A. D. DiChiara, E. Sistrunk, P. Agostini, L. F. DiMauro, D. A. Reis, "Observation of high-order harmonic generation in a bulk crystal" Nature Phys. **7**, 138-141 (2011).

[3] Y. S. You, D. A. Reis, S. Ghimire, "Anisotropic high-harmonic generation in bulk crystals" Nature Phys. **13**, 345-350 (2017).

[4] M. R. Bionta, E. Haddad, A. Leblanc, V. Gruson, P. Lassonde, H. Ibrahim, J. Chaillou, N. Émond, M. R. Otto, A. Jiménez-Galán, R. E. F. Silva, M. Ivanov, B. J. Siwick, M. Chaker, F. Légaré, "Tracking ultrafast solid-state dynamics using high harmonic spectroscopy," Phys. Rev. Res. **3**, 023250 (2021).

[5] G. Vampa, T. J. Hammond, N. Thiré, B. E. Schmidt, F. Légaré, C. R. McDonald, T. Brabec, D. D. Klug, P. B. Corkum, "All-optical reconstruction of crystal band structure", Phys. Rev. Lett. **115**, 193603 (2015).

[6] A.A. Lanin, E.A. Stepanov, A.B. Fedotov, A.M. Zheltikov, "Mapping the electron band structure by intraband high-harmonic generation in solids", Optica **4**, 516 (2017).

[7] N. Tancogne-Dejean, O.D. Mücke, F.X. Kärtner, A. Rubio, "Impact of electronic band structure in high-harmonic generation spectra of solids", Phys. Rev. Lett. **118**, 087403 (2017).

[8] J. Park, A. Subramani, S. Kim, M. F. Ciappina, "Recent trends in high-order harmonic generation in solids", Advances in Physics: X, 7:1, 2003244 (2022).

[9] H. Lakhotia, H. Y. Kim, M. Zhan, S. Hu, S. Meng, E. Goulielmakis, "Laser picoscopy of valence electrons in solids", Nature **583**, 55 (2020).

[10] G. Vampa, T. Brabec, "Merge of high harmonic generation from gases and solids and its implications for attosecond science", J. Phys. B **50**, 083001 (2017).

[11] K. Nakagawa, H. Hirori, S. A. Sato, H. Tahara, F. Sekiguchi, G. Yumoto, M. Saruyama, R. Sato, T. Teranishi, Y. Kanemitsu, "Size-controlled quantum dots reveal the impact of intraband transitions on high-order harmonic generation in solids", Nature Phys. **18**, 874 (2022).

[12] L. Carbone, C. Nobile, M. De Giorgi, F. Della Sala, G. Morello, P. Pompa, M. Hytch, E. Snoeck, A. Fiore, I. R. Franchini, M. Nadasan, A. F. Silvestre, L. Chiodo, S. Kudera, R. Cingolani, R. Krahne, L. Manna, "Synthesis and micrometer-scale assembly of colloidal CdSe/CdS nanorods prepared by a seeded growth approach", Nano Lett. **7**, 2942 (2007).

[13] A. Schleusener, M. Micheel, S. Benndorf, M. Rettenmayr, W. Weigand, M. Wächtler, "Ultrafast Electron Transfer from CdSe Quantum Dots to an [FeFe]-Hydrogenase Mimic", J. Phys. Chem. Lett. **12**, 4385 (2021).

[14] R. Koole, E. Gröneveld, D. Vanmaekelbergh, A. Meijerink, C. de Mello Donegá, "Size effects on semiconductor nanoparticles", In: de Mello Donegá, C. (eds) Nanoparticles. Springer, Berlin, Heidelberg. https://doi.org/10.1007/978-3-662-44823-6_2 (2014).

[15] A. L. Abdelhady, M. Afzaal, M. Azad Malik, P. O'Brien, "Flow reactor synthesis of CdSe, CdS, CdSe/CdS and CdSeS nanoparticles from single molecular precursor(s)", Mater. Chem. **21**, 18768 (2011).

[16] W. W. Yu, L. Qu, W. Guo, X. Peng, "Experimental determination of the extinction coefficient of CdTe, CdSe, and CdS Nanocrystals", Chem. Mater. **15**, 2854 (2003).

[17] A. I. Ekimov, F. Hache, M. C. Schanne-Klein, D. Ricard, C. Flytzanis, I. A. Kudryavtsev, T. V. Yazeva, A. V. Rodina, Al. L. Efros, "Absorption and intensity-dependent photoluminescence measurements on CdSe quantum dots: assignment of the first electronic transitions", J. Opt. Soc. Am. B **10**, 100 (1993).

[18] T. S. Shyju, S. Anandhi, R. Indirajith, R. Gopalakrishnan, "Effects of annealing on cadmium selenide nanocrystalline thin films prepared by chemical bath deposition", Journ. Alloys and Compounds **506**, 892 (2010).

[19] N. Tancogne-Dejean, O.D. Mücke, F.X. Kärtner, A. Rubio, "Ellipticity dependence of high-harmonic generation in solids originating from coupled intraband and interband dynamics", Nature Com. doi: 10.1038/s41467-017-00764-5 (2017).

[20] S. Grimme, C. Bannwarth, P. Shushkov, "A robust and accurate tight-binding quantum chemical method for structures, vibrational frequencies, and noncovalent interactions of large molecular systems parametrized for all spd-block elements (Z = 1–86)", ACS J. Chem. Theory Comput. **13**, 1989 (2017).

[21] N. Tancogne-Dejean, M. J. T. Oliveira, X. Andrade, H. Appel, C. H. Borca, G. Le Breton, F. Buchholz, A. Castro, S. Corni, A. A. Correa, U. De Giovannini, A. Delgado, F. G. Eich, J. Flick, G. Gil, A. Gomez, N. Helbig, H. Hübener, R. Jestädt, J. Jornet-Somoza, A. H. Larsen, I. V. Lebedeva, M. Lüders, M. A. L. Marques, S. T. Ohlmann, S. Pipolo, M. Rampp, C. A. Rozzi, D. A. Strubbe, S. A. Sato, C. Schäfer, I. Theophilou, A. Welden, A. Rubio, "Octopus, a computational framework for exploring light-driven phenomena and quantum dynamics in extended and finite systems", The Journal of Chem. Phys. **152**, 124119 (2020).

[22] N. J. Shevchik, J. Tejeda, M. Cardona, and D. W. Langer, "Photoemission and density of valence states of II-VI compounds: ZnTe, CdSe, CdTe, HgSe and HgTe", Phys. Status Solidi B **59**, 87 (1973).

[23] P. B. Corkum, "Plasma perspective on strong field multiphoton ionization", Phys. Rev. Lett. **71**, 1994 (1993).

[24] G. Vampa, C. McDonald, A. Fraser, T. Brabec, "High-harmonic generation in solids: bridging the gap between attosecond science and condensed matter physics", IEEE Journ. Sel. Top. Quant. Electr. **21**, 8700110 (2015).

[25] G. Vampa, B. Ghamsari, S. S. Mousavi, T. Hammond, A. Olivieri, E. Lisicka-Skrek, A. Y. Naumov, D. Villeneuve, A. Staudte, P. Berini, P. Corkum, "Plasmon-enhanced high-harmonic generation from silicon", Nat. Phys. **13**, 659–662 (2017).

[26] G. Siroki, P. D. Haynes, D. K. K. Lee, V. Giannini, "Protection of surface states in topological nanoparticles", Phys. Rev. Mat. **1**, 024201 (2017).

[27] G. Zograf, K.l Koshelev, A. Zalogina, V. Korolev, R. Hollinger, D.-Y. Choi, M. Zuerch, C. Spielmann, B. Luther-Davies, D. Kartashov, S. V. Makarov, S. S. Kruk, Y. Kivshar, "High-harmonic generation from resonant dielectric metasurfaces empowered by bound states in the continuum", ACS Phot. **9**, 567 (2022).

# Supplemental Material

# Tracing spatial confinement in semiconductor quantum dots by high-order harmonic generation

H. N. Gopalakrishna[1,2], R. Baruah[3,4], C. Hünecke[4], V. Korolev[1], M. Thümmler[4], A. Croy[4], M. Richter[4], F. Yahyaei[5], R. Hollinger[1], V. Shumakova[6], I. Uschmann[1,2], H. Marschner[1,2], M. Zürch[1,7,8], C. Reichardt[3], A. Undisz[9, 10], J. Dellith[3], A. Pugžlys[6], A. Baltuška[6], C. Spielmann[1,2,11], U. Peschel[5,11], S. Gräfe[4,11,12], M. Wächtler[3,4,11], and D. Kartashov[1,11]

1. Institute of Optics and Quantum Electronics, Friedrich-Schiller University Jena, Max-Wien-Platz 1, 07743 Jena, Germany

2. Helmholtz-Institut Jena, Helmholtzweg 4, 07743 Jena, Germany

3. Leibniz Institute of Photonic Technology, Albert-Einstein-Straße 9 07745 Jena, Germany

4. Institute of Physical Chemistry, Friedrich Schiller University Jena, Helmholtzweg 4, 07743 Jena, Germany

5. Institute of Condensed Matter Theory and Solid-State Optics, Friedrich Schiller University Jena, Fröbelstieg 1, 07743 Jena, Germany

6. Institute for Photonics, Vienna University of Technology, Gußhausstrasse. 25-29, 1040 Vienna, Austria

7. Department of Chemistry, University of California, Berkeley, California 94720, USA

8. Materials Sciences Division, Lawrence Berkeley National Laboratory, Berkeley, California 94720, USA

9. Institute of Materials Science and Engineering, Chemnitz University of Technology, Erfenschlager Str. 73, 09125 Chemnitz, Germany

10. Otto Schott Institute of Materials Research, Friedrich Schiller University Jena, Löbdergraben 32, 07743 Jena, Germany

11. Abbe Center of Photonics, Albert-Einstein-Straße 6, 07745 Jena, Germany

12. Fraunhofer Institute for Applied Optics and Precision Engineering, Albert-Einstein-Str.7, 07745 Jena, Germany

**EXPERIMENTAL**

*Chemicals*

Trioctylphosphine oxide (TOPO, 99 %), Trioctylphosphine (TOP, 97%), Cadmium oxide (CdO, 99.99 %), Selenium (Se, 99.99 %), Toluene ( 99.8 % anhydrous) and Methanol (99.8 % anhydrous) were purchased from Sigma Aldrich and Octadecylphosphonic acid (ODPA, 97 %) was purchased from Carl Roth.

*Synthesis of colloidal CdSe quantum dots*

Colloidal CdSe qunatum dots were prepared via hot injection as described in Ref.[Carbone2007, Schleusener21]. In a 25 mL three neck flask with a magnetic stirrer inside, 60 mg CdO, 0.28 g ODPA and 3 g TOPO were mixed, heated up to 150°C and evacuated for 1 hour. Then, under nitrogen flow the solution was heated up which turned clear and colourless at 310°C. At this point 1.5 g TOP was injected into the flask and the solution was allowed to heat up again. A TOP:Se solution (0.058 g TOP + 0.36 g Se) was injected into the solution at varying the injection temperature and the growth time was chosen according to the desired size of the quantum dot. Size 2.2 nm diameter was achieved by injecting the TOP:Se solution at 330°C and cooling down the solution by immediately removing the heating mantle and using cold water around the flask for faster cooling. Larger paricles of sizes 4.4 nm, 6 nm and 8 nm diameter were achieved by injecting the TOP:Se solution at 320°C and extending the growth period to 1.5 min, 5 min and 1 hour respectively at 300°C followed by cooling down the flask. When the temparature was around 80°C, 10 mL of toluene was injected. After the synthesis, the particles were cleaned by precipitation with MeOH and centrifugation at 5300 RPM for 10 min, followed by discarding the supernantant and redispersing the QD solid in 10 mL toluene. The cleaning procedure was performed four times in total. The final product dispersed in 10 mL toluene and stored under inert atmosphere.

*Deposition of colloidal quantum dots into layers*

Prior to deposition of colloidal QDs, 1 cm x 1 cm sapphire substrates (purchased from CrysTec) were cleaned with acetone, methanol, hellmanex and water with 20 min ultrasonication each followed by air drying. 200 µL of each 2.2 nm QD (0.44 mM), 3 nm QD (0.025 mM), 4 nm QD (6.8 µM) and 8.2 nm QD (2.8 µM) solution (in toluene) were drop casted and finally dried under vacuum.

*Preparation of bulk CdSe layer*

The Bulk CdSe layer was prepared via chemical bath deposition as reported by Shyju et al. [2]. In the first step $Na_2SeSO_3$ was prepared. Therefore, in a 25 mL three neck round bottom flask, 110 mg of Se and 40 mg of $Na_2SO_3$ were mixed followed by addition of 7 mL distilled water and stirred for 24 hours at 90°C. Sapphire substrates were cleaned with acetone and isopropanol for 20 min in an ultrasonic bath followed by drying by air blowing. In the second step, in a 25 mL two neck round bottom flask with a magnetic stirrer inside, 620 mg of Cadmium acetate hydrate was mixed with 5 mL distilled water and stirred. 292 mg ethylenediaminetetraacetic acid (EDTA) dissolved in 4 mL ammonia was added. The pH of the solution was adjusted to 10-11 by addition of more $NH_3$. The cleaned sapphire substrate was placed inside the flask with a holder from the top and the solution was heated to 90°C followed by addition of 5 mL $Na_2SeSO_3$. The colour of the solution was observed to be turning into yellow then orange and finally red. Then the solution was stirred for 1 hour more at 90°C. Finally, the solution was removed, and the substrates were washed with distilled water in a Petri dish. In the third step, the substrates were then taken for annealing for 4 hours at 600°C under nitrogen flow to convert the original zinc blende phase to wurtzite phase.

*TEM*

Colloidal QDs were deposited on carbon coated Cu grid (purchased from PLANO GmbH) followed by vacuum drying. Transmission electron microscopy (TEM) - TEM images were recorded in conventional (TEM) and scanning mode (STEM) using a JEM-ARM200F NEOARM (Jeol) operating at 80 kV, equipped with spherical aberration corrector, bright field (BF), annular bright field (ABF) and annular dark field (ADF) detectors. Images were processed using ImageJ 1.53a program [3]. Size and size distribution were evaluated using 500 to 1000 particles for each QD batch.

*SEM*

Scanning electron microscopy images were collected of the layers deposited on silicon wafer using a dual beam Helios NanoLab G3 UC (FEI).

*AFM*

A Bruker Dimension Edge was used in tapping mode to collect the micrographs. To determine film thickness, a scratch was made on the layer and AFM micrographs were collected crossing the scratch.

*X-ray diffraction analysis (XRD)*

Grazing incidence X-ray diffraction (GIXRD) was performed using a Panalytical Xpert Pro MPD (Cu-K$\alpha$1,2) with omega angle 0.4° to 3°. XRD analysis was performed for the bulk CdSe layer and the CdSe QDs. For this, colloidal QDs were deposited on glass substrates via drop casting.

*Steady-state UV/Vis Absorption Spectroscopy*

UV/Vis absorption spectra were recorded using a JASCO V780 UV-visible/NIR spectrometer for QDs dispersed in toluene (3 mL) in a quartz cuvette with 1 cm path length. Similarly, UV/vis absorption spectra of solid layers were recorded using a sample holder suitable for 1 cm x 1 cm size layers. Absorption spectra of the bulk CdSe layer (deposited on sapphire) was recorded using a clean sapphire substrate as reference.

*Photoluminescence Spectroscopy*

Photoluminescence spectra of the quantum dots dispersed in toluene were collected by exciting at 3.1 eV (400 nm) using a FLS980 photoluminescence spectrometer (Edinburgh Instruments Ltd. Livingston, UK). Absolute photoluminescence yield $\Phi_{PL}$ determination was performed in an integrating sphere set up with a 1 cm path length quartz cuvette filled with QDs dispersed in toluene with around 0.1 optical density. Photoluminescence spectra of the quantum dot layers were collected by exciting at 3.1 eV (400 nm) using a Horiba Scientific Fluorolog-3 spectrofluorometer. Spectra collection was done in a so-called front face mode, placing the front face of the sample in 22.5 ° to the excitation beam.

**Optical properties of QDs**

In the absorption spectra of the quantum dots, characteristic electronic transitions are observed which can be assigned to the lowest band edge $1S_e$-$2S_{3/2(h)}$ and to the higher energy $1P_e$-$1P_{3/2(h)}$ transitions. With increasing size, the position of the first excitonic peak is redshifted (Figure S1). The diameter of the QDs was estimated from the spectral position of first excitonic transition in the absorption spectra by the method described by Peng et al. [4] and compared to results from TEM images. The results from both methods are in good agreement, deviations for the largest particles are caused by a slight anisotropy of the particles and the uncertainty of the exact peak position due to broad spectral features. Similar to the peak position in absorption spectra, the band edge luminescence peak is redshifted with increasing QD size. Additionally, for the smallest particles a broad emission band at wavelength longer than 600 nm is observed originating from trap state emission (Figure S1). Surface trap state emission appears more pronounced in smaller QDs due to a higher surface to volume ratio as compared to larger QDs. The quantum yield of the band edge luminescence decreases with increasing particle size in accordance with report in literature [5].

The band gap ($E_g$) of QDs with different sizes was derived from their absorption spectra via Tauc plots and compared to the position of the band edge photoluminescene band position (Fig.S2). The results from both methods show good agreement (Table S1).

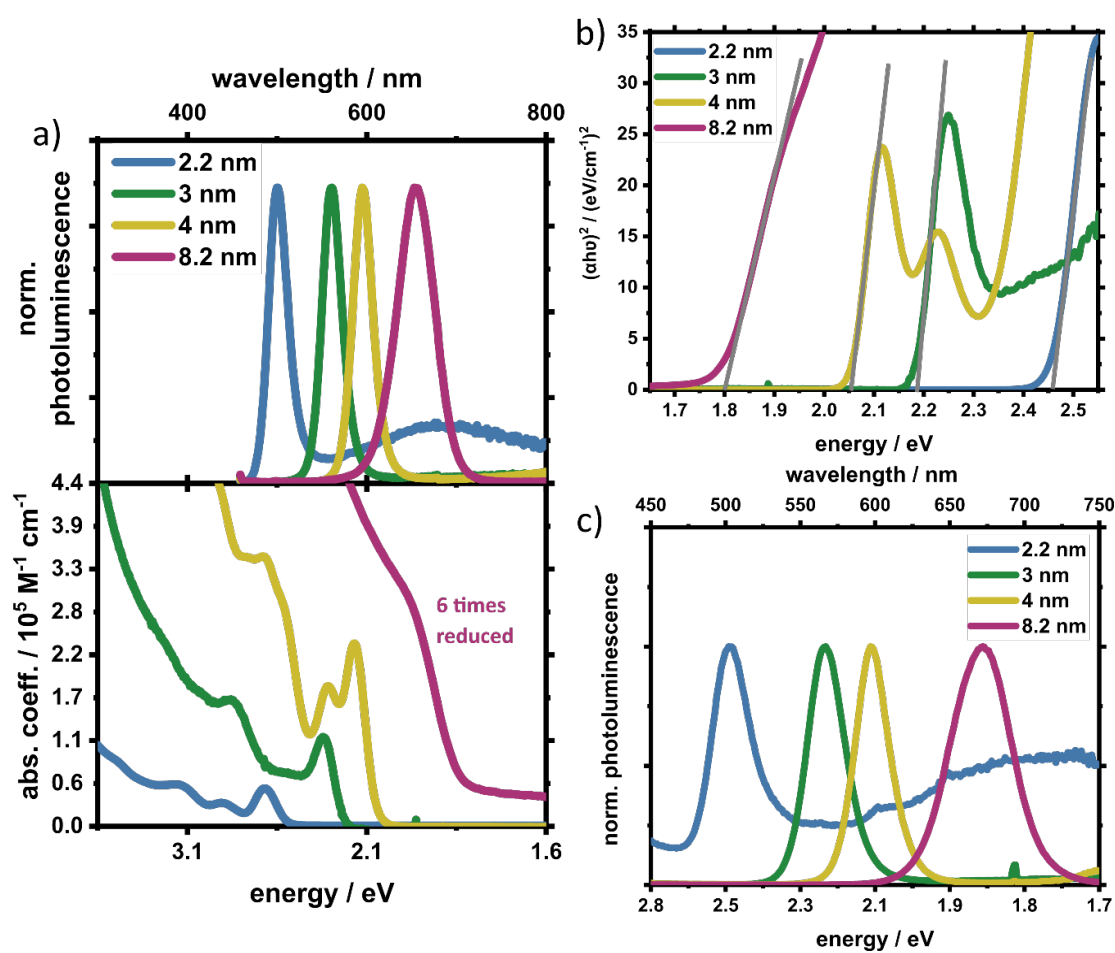

**Fig. S1.** a) UV/vis absorption (bottom panel) and normalized photoluminescence spectra (top panel) of CdSe QDs of different sizes used in this investigation dispersed in toluene, b) Tauc plots for different sizes of CdSe QDs derived from absorption spectra of particles dispersed in toluene and c) normalized photoluminescence spectra of different sizes of CdSe QDs deposited in layer.

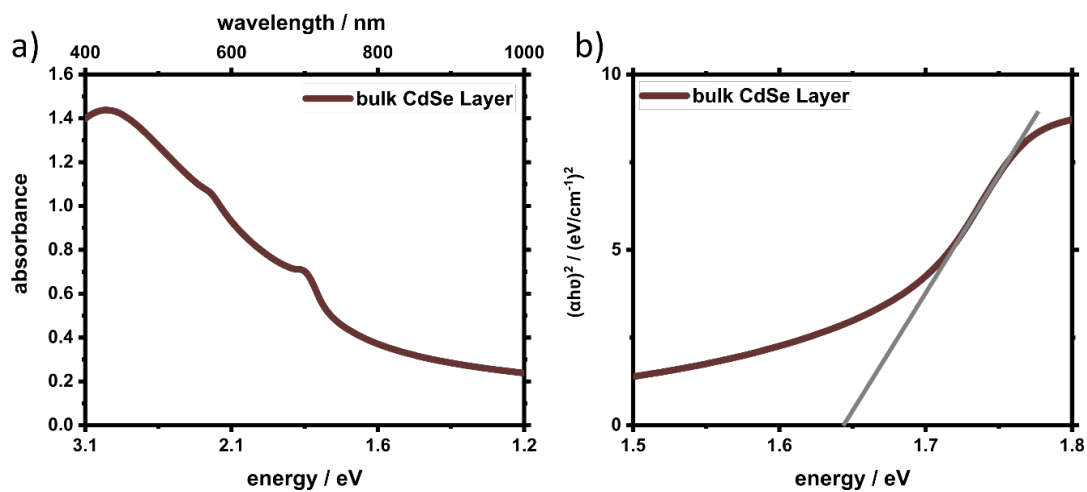

**Fig. S2.** (A) UV/vis absorption spectra and (B) Tauc's plot of the bulk CdSe layer.

**Table S1.** Sizes, spectral characteristics, quantum yields of the photoluminescence $\Phi_{PL}$ and band gaps of different CdSe QDs presented as different colors as in Figure 1.

| Size (Absorption spectra) / nm | 1$^{st}$ absorption peak / eV | Band edge photoluminescence peak / eV | Band Gap ($E_g$)/ eV (Tauc plot) | $\Phi_{PL}$ / % | Size (Absorption spectra) / nm |
|---|---|---|---|---|---|
| 2.2 | 2.55 | 2.48 | 2.48 ± 0.05 | 11 ± 2 | 2.2 |
| 3 | 2.25 | 2.21 | 2.18 ± 0.07 | 2.6 ± 2 | 3 |
| 4 | 2.11 | 2.08 | 2.05 ± 0.06 | 1.5 ± 2 | 4 |
| 8.2 | 1.9 | 1.89 | 1.89 ± 0.02 | 0.3 ± 2 | 8.2 |

**XRD – quantum dots**

XRD pattern in Figure S3 corresponds to pure wurtzite CdSe with the characteristic reflection from (010), (011), (110), (013) and (112) at 23.93°, 25.38°, 42.0°, 45.84° and 49.74° (ICSD:60630). As an indication of smaller particles relatively broad XRD peaks were observed for small quantum dots.

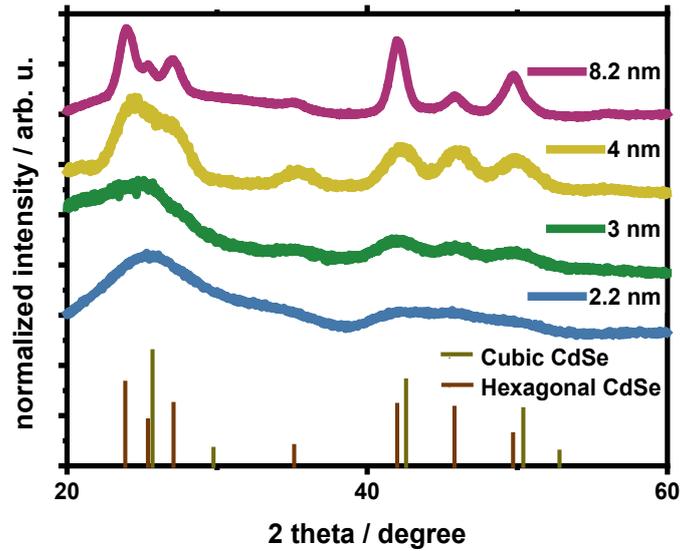

**Fig. S3.** XRD patterns of CdSe QDs of different sizes.

**XRD – bulk sample**

The XRD of the bulk CdSe layer clearly shows that the sample has wurtzite phase. The hkl indices match very well the reference data for hexagonal wurtzite CdSe. From the fact, that we see all the tabulated diffraction peaks and the intensity distribution correlates with the structure factor, we must expect no strong preferred crystal orientation. From the FWHM of the reflexes the crystallite size was determined using the Scherrer formula (program Highscore+) showing that the layer consists of crystallites with sizes in the order of 50 nm. The diffraction profiles are quite narrow and homogeneous. There is no indication of much different crystallite sizes. Further, one can expect no strong crystallites defects which would affect the symmetry of the profiles.

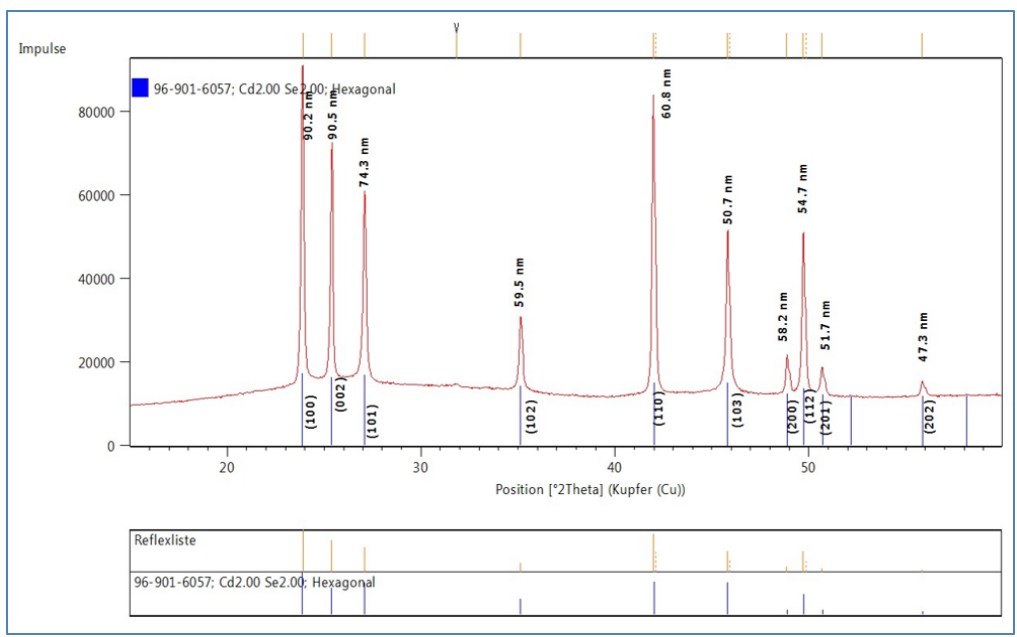

**Fig. S4.** XRD of bulk CdSe layer.

## Transmission electron microscopy (TEM)

TEM images were used to determine the average sizes and size distribution of QDs as depicted in Figure S5 and Table S2. Aspect ratio of the particle which is not perfectly spherical were also determined.

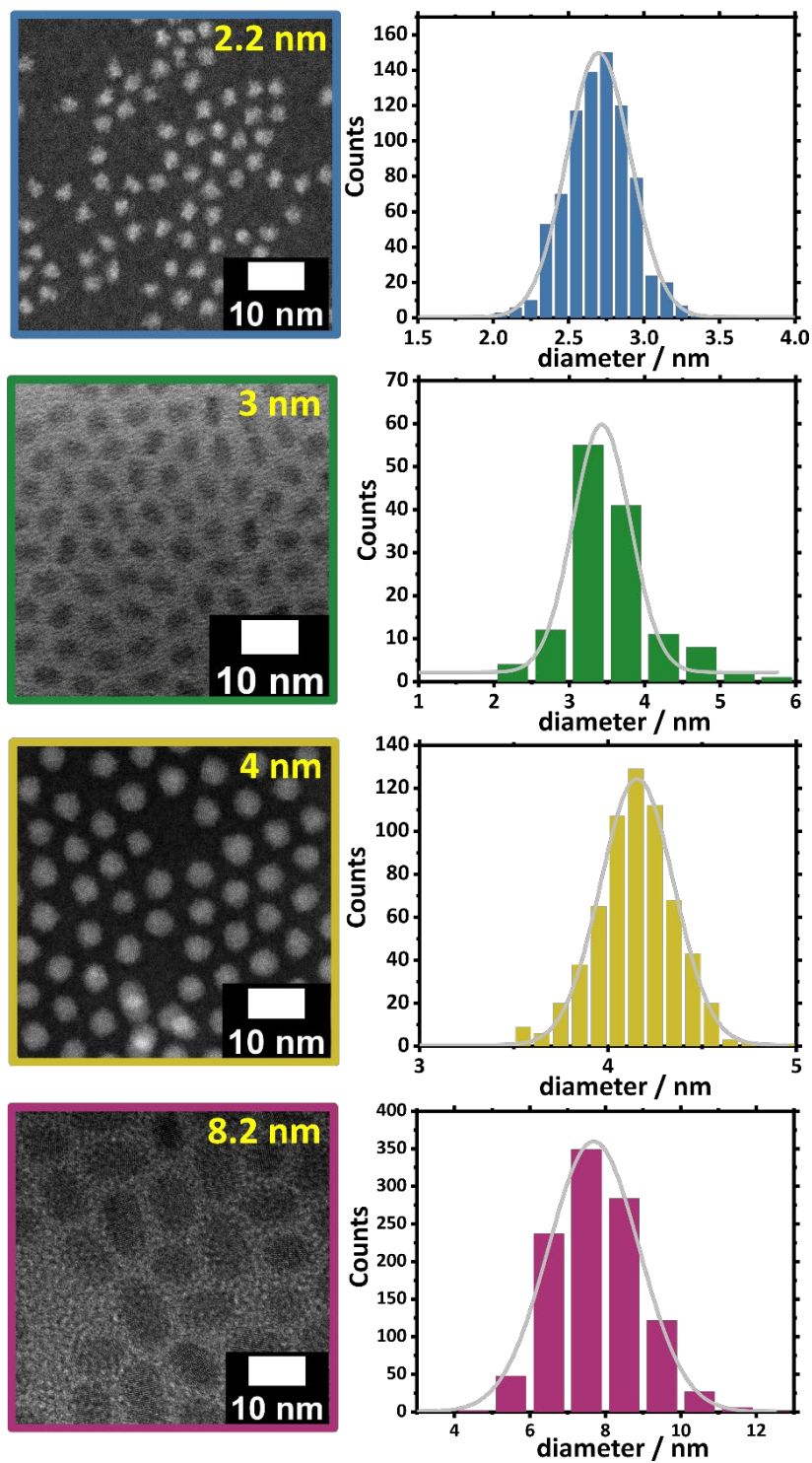

**Fig. S5.** TEM images and size distribution of different sizes of CdSe QDs shown by blue, green, yellow and red colour (border of the images and histograms).

**Table S2.** Size, size distribution and aspect ratio of different sizes of CdSe QDs calculated from the TEM images in Figure S5.

| Size (Absorption spectra) / nm | Size (TEM) / nm | FWHM /nm | Aspect ratio |
|---|---|---|---|
| 2.2 | 2.6 | 0.5 | - |
| 3 | 3.4 | 0.89 | - |
| 4 | 4.2 | 0.38 | - |
| 8.2 | 7.7 | 2.82 | 1.7 : 1 |

**Scanning electron microscopy (SEM)**

Scanning electron microscopy images were collected of the layers deposited in Silicon wafer using a dual beam Helios NanoLab G3 UC (FEI). As shown in Figure S6, organized array of quantum dots was observed. Although SEM images were collected in similar imaging conditions, due to different amount of organic residuals left on QD batches, there are differences in the contrast and image quality.

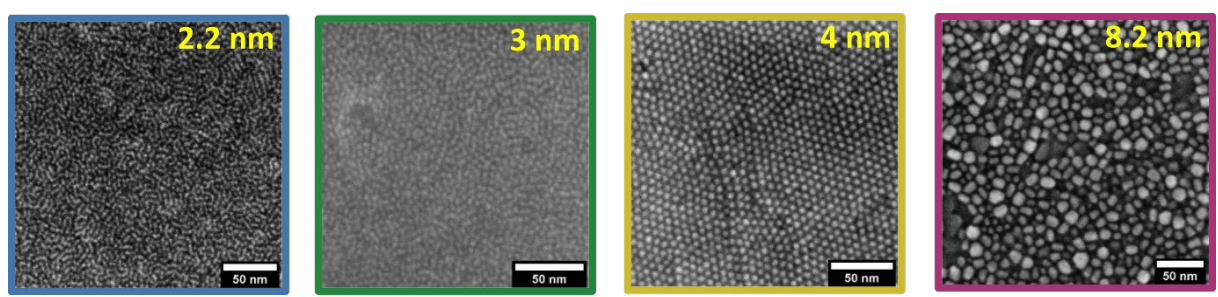

**Fig. S6.** SEM images of CdSe QDs of sizes 2.2 nm (blue), 3 nm (green), 4 nm (yellow) and 8.2 nm (red). 2 and 3 nm sized QDs are below or just at the limit of the resolution of the instrument.

**Atomic force microscopy (AFM)**

Atomic force microscopy technique was used to determine the thickness of the layers of both bulk and Quantum dots as depicted in Figure S7 & Table S3. For this a scratch was made on the QD layer and AFM micrographs were collected over the scratch. Thickness was determined from the thin film surface to the bottom of the scratch as shown in Figure S7 (B). A mean value was determined in both the surface and bottom scratch region of the graph (Figure S7 (B)) and subtracted to determine the thickness.

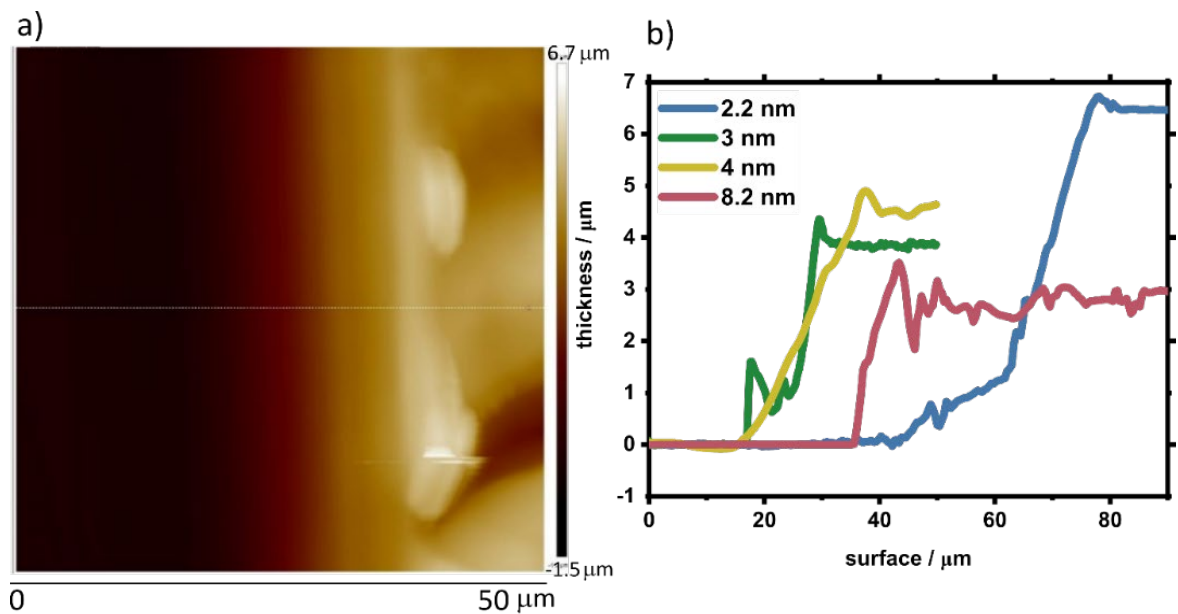

**Fig. S7.** a) AFM image of 4 nm QD. b) Thickness determination curve of the QD layers, where thickness is calculated from the bottom of scratch (at 0) to the top of the surface.

**Table S3.** Thickness of the CdSe QD and bulk CdSe layers determined from atomic force micrographs.

| Size (Absorption spectra) / nm | Thickness / μm |
|---|---|
| 2.2 | 6.4 ± 0.1 |
| 3 | 3.8 ± 0.1 |
| 4 | 4.5 ± 0.12 |
| 8.2 | 2.8 ± 0.24 |
| Bulk CdSe | 0.140 ± 0.012 |

**Intensity dependence of HHG in QD**

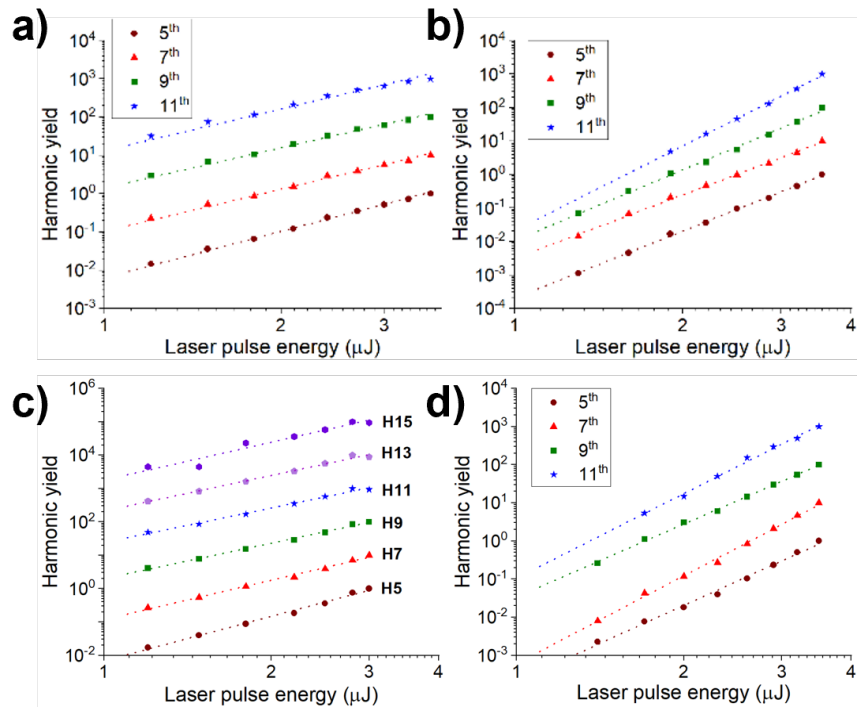

**Fig. S8.** Intensity dependence of harmonic's yield for the 3.72 µm laser wavelength: a) for the 4.4 nm size QD's and b) 2.2 nm size QD's. Intensity dependence of harmonic's yield for the 4.74 µm laser wavelength: c) for the 4.4 nm size QD's and d) 2.2 nm size QD's.

**Ellipticity dependence of HHG in QD**

Polarization dependence of harmonic yield was measured for both 3.75 µm and 4.74 µm laser wavelengths and fixed laser intensity 0.5 TW/cm$^2$. The intensity value was chosen to avoid sample degradation during the data acquisition. Laser polarization was controlled by a broadband (3-6 µm spectral range) achromatic quarter-wave plate (B-Halle), automatically rotated with a step size 1°. HHG spectrum was measured for each waveplate angle, forming a spectrogram that is shown in Fig.S9a for the case of the reference film measurements. The spectrum of each individual harmonic was integrated then in the frequency domain in the range $[2n, 2n + 2]$ where $n$ is the harmonic order, and the correspondent integral value is defined as the yield. The normalized dependence of the yield for individual harmonics as a function of the ellipticity in the laser polarization is shown in Fig.S9c,d for the case of the reference film and 2.2 nm size QD's. The width of these dependencies is defined at half maximum level (FWHM) and the dependence of this width on the laser polarization is shown in the main text in Fig.5. The spectrograms and integral yields for other samples, used in the experiments, look similar to the examples in Fig.S9.

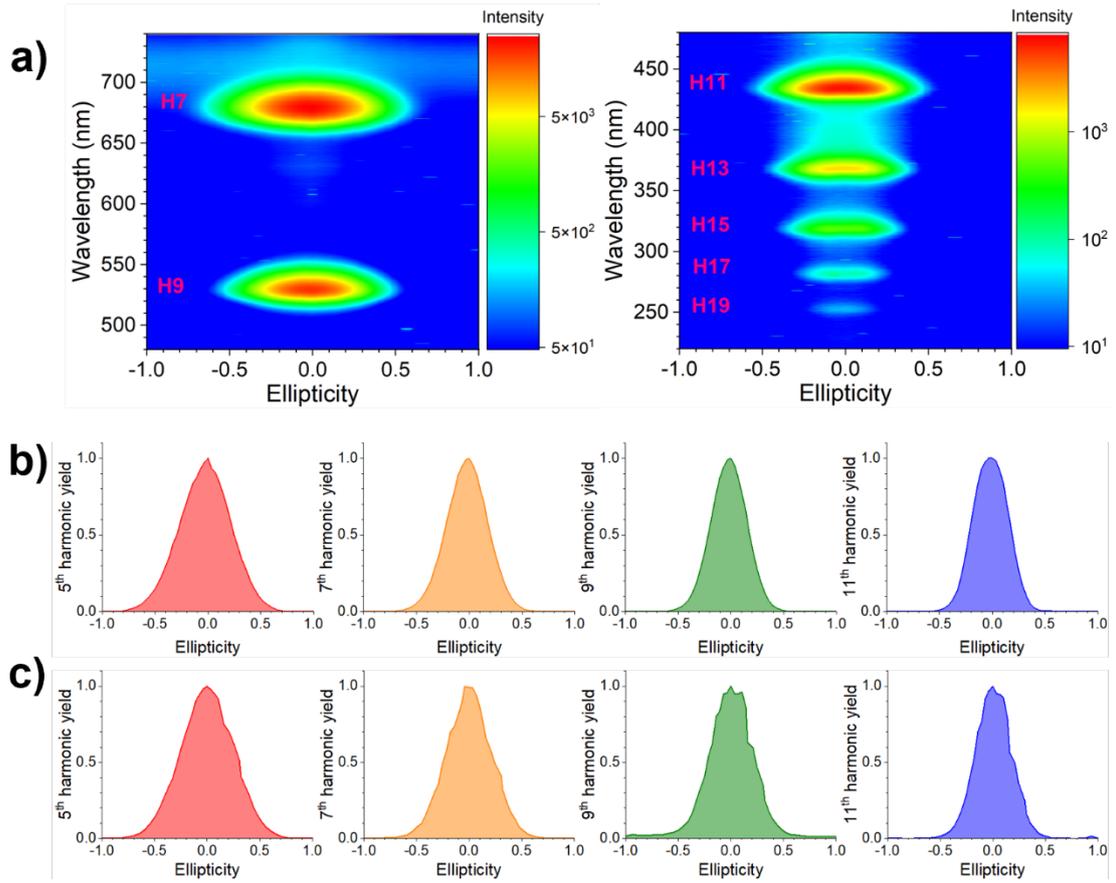

**Fig. S9.** Ellipticity dependence of harmonic's yield. a) A spectrogram for the reference film. Note the logarithmic scale of the spectral intensity. Spectrally integrated harmonic yield for a) the reference film and b) 2.2 nm size QD's.

**NUMERICAL**

**Real-time real-space TDDFT calculations**

The real-time real-space time-dependent density functional theory (rt-TDDFT) simulations were performed using the code Octopus [6]. For the rt-TDDFT simulations, local density approximation (LDA) in the form of Slater-exchange, modified Perdew-Zunger-correlation functionals [7] and LDA-based pseudopotentials from [8] were used.

For the bulk calculations, periodic boundary conditions were used whose k-space was discretized by 32x32x20 k-points. The laser field polarization was oriented along the Γ-K direction, and no rotational averaging was applied because of the very high computational costs of the simulations. The bulk structure had wurtzite CdSe symmetry with lattice parameters a=4.394 Å, b= 7.171 Å [9].

Calculations with CdSe particles of different size ranging from 4 atoms, over 1nm particle size (16 atoms) to 1.5 nm particle size 64 atoms - the largest size that could be simulated within a reasonable time scale - were performed. For these calculations, energy-optimized geometries of dots were used. The radius of the spherical real-space grid was 25, 30 and 35 Bohr (depending on particle size), including 5 Bohr of complex absorbing potential at the edge of the simulation box to avoid wavefunction reflections.

The linearly polarized laser pulse was defined by a $\sin^2$ envelope with an amplitude of $1.8·10^{-3}$ in atomic units, corresponding to an intensity of $1.2·10^{11}$ W/cm$^2$, frequency 0.012 a.u. (3.8 µm wavelength) and a few-cycle duration, limited by the calculation costs (see Fig.S11a). The HHG spectrum was calculated by the Fourier transform of the x-component (along the laser polarization) of the dipole acceleration. The spectra for the bulk and the 64-atom QD are shown in Fig.11b. Clear suppression of harmonics above the 3$^{rd}$ is observed in the dot in comparison to the bulk material. We have to note here, that the calculation results for the bulk may not have fully converged. The reason is that, for good convergence in rt-TDDFT calculations by OCTOPUS, a mesh consisting of at least 70 grid points along each direction in k-space is recommended [10]. This resolution for time-dependent simulations is currently not feasible for the wurtzite bulk CdSe structure. In general, such simulations are computationally extremely demanding and costly and their difficulty sharply increases with increasing the laser wavelength. *Ab initio* calculations for HHG in solids were reported so far either for few-optical-cycle near-IR laser pulses [11] or mid-IR laser pulses but in highly symmetric materials like crystalline Si [12], using one of the world's top large computer clusters. We compared the results of HHG spectra calculations, obtained with the highest possible resolution 32x32x20 k-points grid, with calculations employing 16x16x10 grid and found little difference in harmonic intensities for harmonics up to 15$^{th}$ order. Therefore, we believe, that the 2-3 orders of magnitude difference in harmonic intensities between the bulk and QD spectra shown in Fig.S10 correctly reflects the effect of strong HHG suppression in dots in comparison to the bulk.

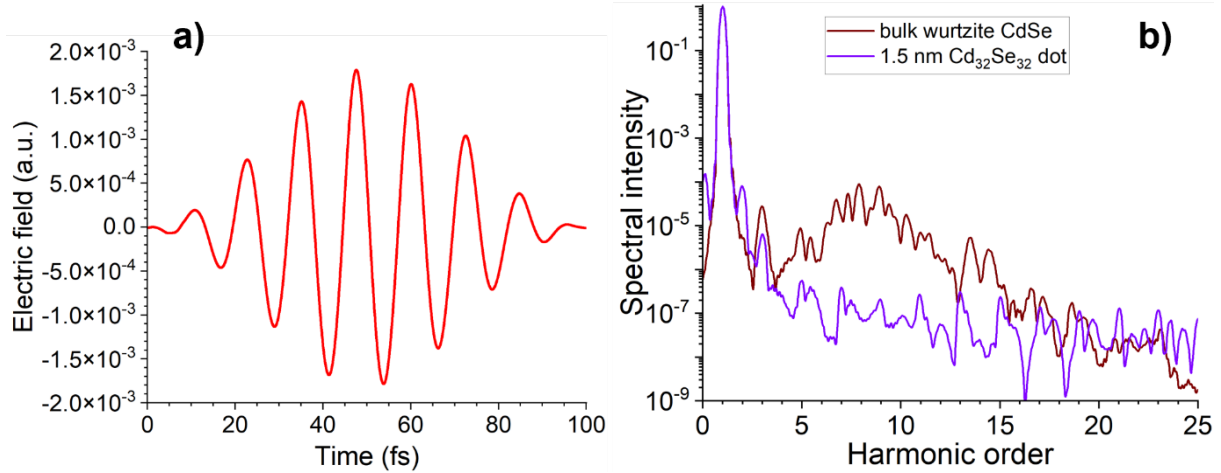

**Fig. S10.** a) Modelled electric field (in atomic units) in the laser pulse. b) HHG spectra for the bulk (wine curve) and 64 atoms QD (violet curve). The spectra are normalized to their maxima.

# Classical Monte-Carlo trajectory simulations

To define the initial probability distribution, the quantum mechanical ground state was calculated using the imaginary time propagation [13] on a grid ranging from $-3d$ to $3d$ with 2048 spatial grid points and a time step size of $\tau = 0.2$ a.u. until the energy was converged. Afterwards, the Wigner-Distribution is calculated. We used rejection sampling in a uniformly sampled rectangle in the phase space to setup the initial conditions for the $10^6$ trajectories. The maximum position and momentum were chosen, such that the remaining quasi-probability adds up to less than 0.01. The laser pulse was modeled as $E(t) = E_0 sin^2\left(t/\tau_p\right) sin\omega t$ with a pulse duration 75 fs FWHM, intensity 0.3 TW/cm², and frequencies corresponding to 3.75 µm or 4.75 µm wavelengths. The field within and outside the dot was calculated, using the analytical solution for the potential of a dielectric sphere exposed to a spatially homogeneous external electric field, i.e. $V = -\frac{3\epsilon_v}{\epsilon_d + 2\epsilon_v} xE(t)$, for $|x| < d/2$ and $V = -xE(t) + \frac{\epsilon_d - \epsilon_v}{\epsilon_d + 2\epsilon_v} \frac{xd^3}{|2x^3|} E(t)$ otherwise with a vacuum permitivity of $\epsilon_v = 1$ and a dot permitivity of $\epsilon_d = 6$ [14]. The dense time propagation of all the trajectories was done using the scipy.integrate.LSODA function.

# References

[1] A Schleusener, M. Micheel, S. Benndorf, M. Rettenmayr, W. Weigand, M. Wächtler, "Ultrafast electron transfer from CdSe quantum dots to an [FeFe]-Hydrogenase Mimic", J. Phys. Chem. Lett. 12, 4385–4391 (2021).

[2] T.S.Shyju, S. Anandhi, R. Indirajith, R. Gopalakrishnan, "Effects of annealing on cadmium selenide nanocrystalline thin films prepared by chemical bath deposition", Journ. Alloys and Compounds 506, 892-897 (2010).

[3] J. Schindelin, I. Arganda-Carreras, E. Frise, V. Kaynig, M. Longair, T. Pietzsch, S. Preibisch, C. Rueden, S. Saalfeld, B. Schmid, J.-Y. Tinevez, D. J. White, V. Hartenstein, K. Eliceiri, P. Tomancak, A. Cardona, "Fiji: an open-source platform for biological-image analysis", Nature Methods 9, 676–682 (2012).

[4] W. W. Yu, L. Qu, W. Guo, X. Peng, "Experimental Determination of the Extinction Coefficient of CdTe, CdSe, and CdS Nanocrystals", Chem. Mater. 15, 2854–2860 (2003).

[5] J. Zhang, X. Zhang, J. Y. Zhang, "Size-Dependent Time-Resolved Photoluminescence of Colloidal CdSe Nanocrystals", J. Phys. Chem. C 113, 9512–9515 (2009).

[6] N. Tancogne-Dejean, M. J. T. Oliveira, X. Andrade, H. Appel, C. H. Borca, G. Le Breton, F. Buchholz, A. Castro, S. Corni, A. A. Correa, U. De Giovannini, A. Delgado, F. G. Eich, J. Flick, G. Gil, A. Gomez, N. Helbig, H. Hübener, R. Jestädt, J. Jornet-Somoza, A. H. Larsen, I. V. Lebedeva, M. Lüders, M. A. L. Marques, S. T. Ohlmann, S. Pipolo, M. Rampp, C. A. Rozzi, D. A. Strubbe, S. A. Sato, C. Schäfer, I.

Theophilou, A. Welden, A. Rubio, "Octopus, a computational framework for exploring light-driven phenomena and quantum dynamics in extended and finite systems", The Journal of Chem. Phys. 152, 124119 (2020).

[7] J. P. Perdew, A. Zunger, "Self-interaction correction to density-functional approximations for many-electron systems", Phys. Rev. B **23**, 5048 (1983).

[8] C. Hartwigsen, S. Goedecker, J. Hutter, "Relativistic separable dual-space Gaussian pseudopotentials from H to Rn", Phys. Rev. B 58, 3641 (1998).

[9] https://materialsproject.org/materials/mp-1070/

[10] N. Tancogne-Dejean, private communications

[11] I. Floss, C. Lemell, G. Wachter, V. Smejkal, S. A. Sato, X.-M. Tong, K. Yabana, J. Burgdörfer, "Ab initio multiscale simulation of high-order harmonic generation in solids", Phys. Rev. A 97, 011410 (2018).

[12] N. Tancogne-Dejean, O. D. Mücke, F. X. Kärtner, A. Rubio, "Impact of electronic band structure in high-harmonic generation spectra of solids." Phys. Rev. Lett. 118, 087403 (2017).

[13] L. Lehtovaara, J. Toivanen, J. Eloranta, "Solution of time-independent Schrödinger equation by the imaginary time propagation method." J. Comput. Phys. 221, 148–157 (2007).

[14] J. D. Jackson, Classical Electrodynamics, Third Edition Eq.4.54 p 184, (John Wiley & Sons, Inc., 1998)